\documentclass[onecolumn]{aastex62}
\usepackage{amsmath, amsthm,amssymb}
\usepackage{natbib}
\usepackage{graphicx, subfigure}
\usepackage{enumitem}
\usepackage{CJK}

\shorttitle{Col-OSSOS: The Colours of OSSOS}
\shortauthors{Schwamb, Fraser, Bannister et al.}

\begin{document}
\begin{CJK*}{UTF8}{bkai}
\title{Col-OSSOS: The Colours of the Outer Solar System Origins Survey} 

\author[0000-0003-4365-1455]{Megan E. Schwamb}
\correspondingauthor{Megan E. Schwamb}
\email{mschwamb.astro@gmail.com}
\affiliation{Gemini Observatory, Northern Operations Center, 670 North A'ohoku Place, Hilo, HI 96720, USA}
\affiliation{Institute of Astronomy and Astrophysics, Academia Sinica; 11F of AS/NTU Astronomy-Mathematics Building, Nr. 1 Roosevelt Rd., Sec. 4, Taipei 10617, Taiwan}

\author[0000-0001-6680-6558]{Wesley C. Fraser}
\affiliation{Astrophysics Research Centre, Queen's University Belfast, Belfast BT7 1NN, UK}

\author[0000-0003-3257-4490]{ Michele T. Bannister}
\affiliation{Astrophysics Research Centre, Queen's University Belfast, Belfast BT7 1NN, UK}
\affiliation{Department of Physics and Astronomy, University of Victoria, Elliott Building, 3800 Finnerty Rd, Victoria, BC V8P 5C2, Canada}
\affiliation{Herzberg Astronomy and Astrophysics Research Centre, National Research Council of Canada, 5071 West Saanich Rd, Victoria, British Columbia V9E 2E7, Canada}

\author[0000-0001-8617-2425]{Micha\"el Marsset}
\affiliation{Astrophysics Research Centre, Queen's University Belfast, Belfast BT7 1NN, UK}
\affiliation{Department of Earth, Atmospheric and Planetary Sciences, MIT, 77 Massachusetts Avenue, Cambridge, MA 02139, USA}

\author[0000-0003-4797-5262]{Rosemary E. Pike}
\affiliation{Institute of Astronomy and Astrophysics, Academia Sinica; 11F of AS/NTU Astronomy-Mathematics Building, Nr. 1 Roosevelt Rd., Sec. 4, Taipei 10617, Taiwan}

\author[0000-0001-7032-5255]{J. J. Kavelaars}
\affiliation{Herzberg Astronomy and Astrophysics Research Centre, National Research Council of Canada, 5071 West Saanich Rd, Victoria, British Columbia V9E 2E7, Canada}
\affiliation{Department of Physics and Astronomy, University of Victoria, Elliott Building, 3800 Finnerty Rd, Victoria, BC V8P 5C2, Canada}

\author[0000-0001-8821-5927]{Susan D. Benecchi}
\affiliation{Planetary Science Institute, 1700 East Fort Lowell, Suite 106, Tucson, AZ, USA, 85719}

\author[0000-0003-4077-0985]{Matthew J. Lehner}
\affiliation{Institute of Astronomy and Astrophysics, Academia Sinica; 11F of AS/NTU Astronomy-Mathematics Building, Nr. 1 Roosevelt Rd., Sec. 4, Taipei 10617, Taiwan}
\affiliation{Department of Physics and Astronomy, University of Pennsylvania, 209 S. 33rd St., Philadelphia, PA 19104, USA}
\affiliation{Harvard-Smithsonian Center for Astrophysics, 60 Garden St., Cambridge, MA 02138, USA}

\author{Shiang-Yu Wang (王祥宇)}
\affiliation{Institute of Astronomy and Astrophysics, Academia Sinica; 11F of AS/NTU Astronomy-Mathematics Building, Nr. 1 Roosevelt Rd., Sec. 4, Taipei 10617, Taiwan}

\author[0000-0002-1506-4248]{Audrey Thirouin}
\affiliation{Lowell Observatory, 1400 W Mars Hill Rd, Flagstaff, AZ 86001, USA}

\author{Audrey Delsanti}
\affiliation{Aix Marseille Universit\'e, CNRS, LAM (Laboratoire d'Astrophysique de Marseille) UMR 7326, 13388, Marseille, France}

\author[0000-0002-6830-476X]{Nuno Peixinho}
\affiliation{CITEUC --- Centre for Earth and Space Science Research of the University of Coimbra, Geophysical and Astronomical Observatory of the University of Coimbra, 3030-004 Coimbra, Portugal}

\author[0000-0001-8736-236X]{Kathryn Volk} 
\affiliation{Lunar and Planetary Laboratory, University of Arizona, 1629 E University Blvd, Tucson, AZ 85721, USA}

\author[0000-0003-4143-8589]{Mike Alexandersen}
\affiliation{Institute of Astronomy and Astrophysics, Academia Sinica; 11F of AS/NTU Astronomy-Mathematics Building, Nr. 1 Roosevelt Rd., Sec. 4, Taipei 10617, Taiwan}

\author[0000-0001-7244-6069]{Ying-Tung Chen (陳英同)}
\affiliation{Institute of Astronomy and Astrophysics, Academia Sinica; 11F of AS/NTU Astronomy-Mathematics Building, Nr. 1 Roosevelt Rd., Sec. 4, Taipei 10617, Taiwan}

\author{Brett Gladman}
\affiliation{Department of Physics and Astronomy, University of British Columbia, Vancouver, BC, Canada}

\author{Stephen D. J. Gwyn}
\affiliation{Herzberg Astronomy and Astrophysics Research Centre, National Research Council of Canada, 5071 West Saanich Rd, Victoria, British Columbia V9E 2E7, Canada}

\author[0000-0003-0407-2266]{Jean-Marc Petit}
\affiliation{Institut UTINAM UMR6213, CNRS, Univ. Bourgogne Franche-Comt\'e, OSU Theta F25000 Besan\c{c}on, France}


\begin{abstract}

The Colours of the Outer Solar System Origins Survey (Col-OSSOS) is acquiring near-simultaneous $g$, $r$, and $J$ photometry of unprecedented precision with the Gemini North Telescope, targeting nearly a hundred trans-Neptunian objects (TNOs) brighter than $m_r=23.6$ mag discovered in the Outer Solar System Origins Survey. Combining the optical and near-infrared photometry with the well-characterized detection efficiency of the Col-OSSOS target sample will provide the first flux-limited compositional dynamical map of the outer Solar System. In this paper, we describe our observing strategy and detail the data reduction processes we employ, including techniques to mitigate the impact of rotational variability.  We present optical and near-infrared colors for 35 TNOs. We find two taxonomic groups for the dynamically excited TNOs, the neutral and red classes, which divide at $g-r \simeq 0.75$. 
Based on simple albedo and orbital distribution assumptions, we find that the neutral class outnumbers the red class, with a ratio of 4:1  and potentially as high as 11:1. Including in our analysis constraints from the cold classical objects, which are known to exhibit unique albedos and $r-z$ colors, we find that within our measurement uncertainty, our observations are consistent with the primordial Solar System protoplanetesimal disk being neutral-class-dominated, with two major compositional divisions in $grJ$ color space.
\end{abstract}

\keywords {Kuiper belt: general  - minor planets, asteroids: general  - planets and satellites: formation - planets and satellites: dynamical evolution and stability  - surveys}


\section{Introduction}

The region beyond Neptune is populated by hundreds of thousands of planetesimals. These trans-Neptunian objects (TNOs)  are the fossils left over after the era of planet construction in our Solar System. Their orbits are a record of the outer Solar System's past evolution (Sections \ref{sec:dynamics}-\ref{sec:history}), and their surface composition is a window into the conditions in the early planet-forming disk (Sections \ref{sec:surfaces}-\ref{sec:motivations}). We developed the Colours of the Outer Solar System Origins Survey (Col-OSSOS) to map the surface properties of the trans-Neptunian populations, through near-simultaneous $g$-, $r$- and $J$-band observations with the 8.1-m Frederick C. Gillett Gemini North Telescope.

Col-OSSOS provides color measurements for a set of TNOs that were detected in a  flux-limited survey with a well-measured detection efficiency. The survey targets TNOs brighter than $m_r = 23.6$ mag found in the Outer Solar System Origins Survey \cite[OSSOS;][]{Bannister2016,Bannister2018}. 
Col-OSSOS affords the first opportunity to explore the true frequency of surface colors within the Kuiper belt, subdivided by dynamical classification.
Initial results are discussed in \citet{Fraser2017,Pike2017} and \citet{Marsset2018}.
We were also able to compare Col-OSSOS TNO measurements to the first interstellar object, 1I/`Oumuamua \citep{Bannister2017}. 

In this paper, we provide an overview of the Col-OSSOS survey, detailing the observing strategy to obtain near-simultaneous optical and near-infrared (NIR) colors (Section \ref{sec:design}) and summarizing our data analysis strategies (Section \ref{sec:colours} and Appendixes~\ref{sec:optical} and ~\ref{sec:NIR}).
Our first data release is 35 TNOs (Section \ref{sec:sample}), which display three color-dynamical taxonomic groups (Section \ref{sec:first_colours}).
We infer the observed and debiased ratio of the two color groups of the red and neutral surfaces within the dynamically excited Kuiper belt population (Section \ref{sec:colour_fraction}).
We briefly examine the implications for the radial color distribution in the primordial planetesimal disk from which the excited TNOs originated (Section \ref{sec:disk}).

\section{The Orbital Structure and Surface Compositions of Trans-Neptunian Objects}

\subsection{Dynamical Populations}
\label{sec:dynamics}


The main Kuiper belt (trans-Neptunian region) can be defined as the conglomeration of minor planets on orbits with semimajor axes between 37 and $\sim50$ au\footnote{More precisely, between the $\nu$18 secular resonance and the 2:1 mean motion resonance with Neptune at 47.5~au, though the cold classicals extend a few au further.}.  The TNO population can be split into two broad dynamical subgroups: the `cold classicals', on near-circular and low-inclination \emph{i} $\lesssim$ 5$^\circ$ orbits, and an overlapping dynamically excited population with \emph{i} $\gtrsim$ 5$^\circ$ \citep{Brown2001, Bernstein2004, Gladman2008, Fuentes2008,Fraser2010,Petit2011,Bannister2018}.
The cold classicals have very little dynamical excitation, with an inclination width of only $\simeq2^{\circ}$  \citep{Brown2001,2008ssbn.book...59K,Gulbis2010, Petit2011}.
In contrast, the dynamically excited Kuiper belt has an inclination width of $\sim$14-16$^{\circ}$ \citep{Petit2017}.

The dynamically excited population is a highly complex structure with several subclasses. 
A fifth of the population is locked in mean motion resonances with Neptune \citep{Malhotra1995,Gladman2012}: their orbits have integer period ratios with Neptune's orbit. 
Non-resonant TNOs with large eccentricities form the scattering disk, a slowly decaying population. 
Their semimajor axes evolve due to active gravitational scattering with Neptune \citep{Gladman2008,Gomes2008}. 
The scattering disk feeds into the short-lived Centaur population, which orbits between the giant planets for timescales of tens of Myr \citep{1999Icar..142..509D,2003AJ....126.3122T}, although it may not be the only population contributing to the Centaurs  \citep{1999AJ....118.1873Y,2010A&A...519A.112D,2010MNRAS.402...13H}.
In contrast, TNOs on $q \gtrsim 38$~au orbits with large $a$ beyond the 2:1 mean motion resonance reside on moderately stable orbits with little or no direct gravitational interactions with Neptune \citep{Emelyanenko2003, Gomes2005,Gladman2008, Brasser2015}. 
Orbits with $a>250$~au and  $q \gtrsim 45$~au are difficult if not impossible for Neptune to directly scatter planetesimals onto during its outward migration \citep{Brasser2015}; these orbits  have unclear origins, and may be emplaced by a different dynamical mechanism \citep{Brown2004, Morbidelli2004, 2006ApJ...643L.135G, 2006Icar..184...59B,2008Icar..197..221K, 2012Icar..217....1B,Trujillo2014,Batygin2016,Sheppard2016,Bannister2017diffusion,2018ApJ...863...45P}. 

\subsection{Migration History and Population Emplacement}
\label{sec:history}

Overwhelming evidence indicates that the Kuiper belt did not form entirely in situ. Planetesimal-driven giant planet migration which scatters the early Solar System's protoplanetesimal disk is the overarching framework \citep{Malhotra1995,Thommes1999,Levison2008,Nesvorny2015b,Nesvorny2015} that best reproduces the observed orbital structure of the trans-Neptunian region. 
The specifics of the dynamical instability result in different signatures in the Kuiper belt's present orbital distribution. 
The dynamically excited population, including most of the resonant objects, were emplaced into the Kuiper belt through gravitational scattering from inward regions of the planetesimal disk. 
Subsequent interactions with Neptune further sculpted the region and are imprinted on the dynamically excited population's present orbital distribution.

The physical properties of the cold classicals -- their photometric colors, slope of their size distribution, albedo distribution, and resolved binary fraction -- significantly differ from the rest of the Kuiper belt \citep[e.g.][]{Tegler2000, Doressoundiram2002, Peixinho2004, Noll2008, Peixinho2008, 2009Icar..201..284B,Fraser2014,Lacerda2014}.
Additionally, the widest cold classical binaries would be disrupted if scattered out into the classical belt by Neptune \citep{Parker2010}. 
Together, this evidence suggests that the vast majority of the cold classicals formed in place. Thus, the cold classicals place unique constraints on Neptune's dynamical history, as their orbits must remain largely undisturbed by Neptune's migration as the planet reached its present-day orbit. 
Recent work by \cite{Fraser2017} using Col-OSSOS measurements shows that this picture is slightly more complicated. Red colored cold classicals are thought to have originated at their present location, but ``blue binaries" (neutral-colored binaries) found within the cold classical belt are thought to be interlopers that formed further inward  and were deposited into the region during the final stages of Neptune migration. 

Recent studies have shown that the smoothness of Neptune's migration at late stages of planetesimal-driven migration produces differences in the predicted structure of Kuiper belt orbits. 
\citet{Nesvorny2015b,Nesvorny2015} showed that Neptune migration with a slow and occasionally jumping ``grainy" motion through a cold disk of planetesimals implants objects on sufficiently excited inclinations and eccentricities in the dynamically excited population.
\citet{Lawler2018sub} find the best match to the near-resonant distant populations is from grainy migration. Work  by \cite{2011ApJ...738...13B}, \cite{2018arXiv180802146R}, and \cite{2018Icar..306..319G}   conclude that  a moderately high eccentric phase during Neptune migration  is also a viable scenario to reproduce the structure of the cold classical belt. 

\subsection{Compositional Surveys}
\label{sec:surfaces}

The past two decades of observations have provided substantial insights into the surface composition of the bright $m_r < 22$ mag TNOs, which are readily studied via optical and NIR  reflectance spectroscopy. 
Their surfaces are divided into three categories: dwarf planets rich in volatile ices such as methane, ethane and water ice \citep{Schaller2007, Barkume2008, Brown2012}, the Haumea collisional family with strong water ice absorption \citep{Brown2007, Schaller2008, Snodgrass2010, Trujillo2011, Carry2012, Fraser2012}, and surfaces devoid of feature-imprinting volatiles other than water ice. 
The majority of spectroscopically studied TNOs are spectrally featureless. 
They exhibit a diversity of surfaces, with spectral gradients ranging from solar-neutral colors to redder than solar in optical wavelengths \citep{2008A&A...487..741A,Barucci2008,2008AJ....135...55B,2011Icar..214..297B}.

For the far more abundant $>$ 22 mag TNOs, optical and infrared spectroscopy is impossible with current ground and space-based facilities.
We must instead rely on what broad-band and narrow-band colors reveal by proxy about the optical and near-infrared spectral slopes. 
Large surveys of TNO surface colors have used a variety of optical and infrared wavelength measurements to attempt to understand and classify TNO surfaces.
\citet{Doressoundiram2008, Peixinho2015}, and the MBOSS database\footnote{\url{http://www.eso.org/~ohainaut/MBOSS/}} \citep{Hainaut2012} provide a compilation of surface colors in the published literature.

The first surveys found that TNO surfaces have a broad range of surface colors, from nearly neutral solar colors to very red \citep{Luu1996}. 
Some dynamical populations are confined to a color range, or exhibit a bimodality in color.  
\cite{2000Natur.407..979T} and \cite{Doressoundiram2001} identified that more neutral surfaces exist at higher orbital inclinations; this trend was later identified as largely an effect of the vast majority of cold classical TNOs having red surfaces \citep{Doressoundiram2001, Tegler2003}. 
Additional work confirmed the statistical significance of the cold classicals' color distribution as  separate from that of other TNOs \citep{Doressoundiram2001,  Doressoundiram2002,Tegler2003, Doressoundiram2007,Peixinho2008}. 
The European Southern Observatory Large Program on Centaurs and TNOs found a continuum of surface colors in the optical and infrared, with relatively linear color slopes from $B$ to $J$ bands \citep{Boenhardt2002, Delsanti2004, Peixinho2004, Delsanti2006}.  
Other photometric surveys identified a bimodality in surface colors of the Centaur population \citep[][]{Peixinho2003,Tegler2003I}. 
As the $r > 22$ mag TNOs were surveyed, a bimodality in color became apparent in smaller-size TNOs \citep{Peixinho2012,Fraser2012,Peixinho2015,Fraser2015}, with a bimodality in $g-i$ in smaller excited TNOs \citep{Wong2017}. 

\subsection{Surface Origins in the Protoplanetesimal Disk}
\label{sec:motivations}

Two distinct models have been put forth to explain the observed diversity of surfaces observed within the small ($<$ 500 km in diameter) TNOs. The two proposals suggest different scenarios for how the surface color variation was emplaced in the TNO population before Neptune migration occurred. Each scenario has its own implications for the structure of the early Solar System's planetesimal disk. Col-OSSOS aims to distinguish between these two ideas.

In the first model, summarized by \cite{DalleOre2013},  small TNO surfaces are divided into five discrete types, each with its own unique color, albedo, and composition. 
The cold classicals fall into their own separate class, while the dynamically excited TNO population is divided into four types. In this scenario, the unique color classes are the direct result of $\sim$5 or more ice lines that existed in the primordial disk: each unique composition corresponds to formation beyond a specific ice line. In the \cite{DalleOre2013} model, these compositional classes were caused by a  sharply striped primordial disk with five or more compositional boundaries and very little or no cross-mixing between the forming protoplanetesimals. In this scenario, individual compositional classes manifest as groups of objects clustered in different regions of the optical-NIR color space; at least 5 should be apparent with sufficiently accurate color measurements.

In the second model, of \citet{Fraser2012}, small TNOs fall into only three compositional classes, which also correlate with dynamical class: the red cold classicals, the neutral dynamically excited objects, and red dynamically excited objects. \cite{Fraser2012} suggest that the protoplanetesimal disk did not show a primordial compositional gradient, but rather, was compositionally homogenous between 15 and 45 au. All planetesimals in this region would have had relatively similar abundances of surface volatiles after formation. The observed compositional classes were thus a result of post-formation evolution, with some objects rapidly losing their light volatile species. What volatile was lost depended only on surface temperature and hence formation location, with each class of TNO predominantly residing at a different distance within the protoplanetesimal disk \citep{Wong2016,Wong2017}. The separate long-term chemical evolution pathways resulted in different surface colors dividing into 3 broad classes (the red cold classicals, the neutral dynamically excited objects, and red dynamically excited objects). In the \cite{Fraser2012} scenario, the range of surface colors seen in each class is the result of a range of mixing of unique surface materials. Only two taxons will be apparent in the optical and NIR space, each exhibiting a range of optical-NIR colors; with the cold classical objects sharing a taxon with the dynamically excited red objects.

A  sample of sufficiently accurate optical and NIR TNO colors should be able to differentiate between the scenarios of \cite{DalleOre2013} and \cite{Fraser2012} and inform us of the compositional properties of the early planetesimal disk. Combining the dynamics of the ensemble Kuiper belt together with its physical and chemical properties would create a powerful probe of Neptune's migration and of the compositional structure of the primordial disk from which the TNOs originated. 
To date, this has proven a challenging task. 
Most TNO physical property studies examine the hodgepodge set of objects that were discovered by various surveys with different and varying detection biases.
Object size and dynamical classification are dependent properties: small TNOs become discoverable when they are near perihelion, at closer heliocentric distances. 
For example, Centaurs with color measurements are on average much smaller in size than the hot classical objects with measured colors.
This has made it difficult, if not impossible, to accurately estimate the true frequency of the different surface color groups in the modern-day Kuiper belt.
Thus, a careful sample of TNOs with known discovery biases is necessary to disentangle the effects of observational biases from the color distribution of the intrinsic populations \citep{pike2013}.

\section{Survey Design}
\label{sec:design}

The goal of Col-OSSOS is to produce the first ever flux-limited optical and near-infrared color survey of TNOs, with well-characterized and well-quantified biases. 
We aim to use this sample to test and probe the taxonomic classes within the Kuiper belt and to produce reliable intrinsic population statistics. 
The Col-OSSOS program is governed by a set of overarching requirements:
\begin{itemize}[itemsep=-3pt]
\item a well understood, flux-limited TNO sample with quantified biases (Section \ref{sec:targets})
\item color measurements that can distinguish between the disk models of \citet{DalleOre2013} and \citet{Fraser2012} (Section \ref{sec:filters})
\item homogeneity in our observing scheme (Section \ref{sec:observing_strategy})
\item observations in different filters acquired as temporally close as possible (Section \ref{sec:observing_strategy})
\item consistent high photometric quality of the observations for all targets (Section \ref{sec:observing_strategy}).
\end{itemize}

\subsection{Target TNO Selection}
\label{sec:targets}

All Col-OSSOS targets are drawn from the sample of over 800 TNOs and Centaurs  ($m_r = 21.8$-25.2 mag; $3.6 < H_r < 14.5$ mag)  found by the Outer Solar System Origins Survey \citep[OSSOS:][]{Bannister2016,Bannister2018}, a survey with well-characterized survey biases.
OSSOS was a wide-field $r$-band survey with the 3.58-m Canada-France-Hawaii Telescope's (CFHT) MegaPrime square-degree field-of-view (FOV) imager (MegaCam) \citep{Boulade2003}. 
Operating from 2013-2017, OSSOS searched eight $20$ deg$^2$ regions (blocks) of sky \citep{Bannister2018}, at locations where resonant TNOs come to perihelion and become optimally detectable. 
Strong emphasis was placed on thorough recovery and tracking of discoveries in a dense observing cadence across two years, freeing the discovered TNOs from the challenges of ephemeris bias \citep{Jones2006}.  97\% of the OSSOS TNO sample have fractional semimajor axis uncertainty of $\sigma_a < 0.1$\% \citep{Bannister2016,Bannister2018}. 
The biases that surveying imposed on the TNO sample are well-quantified for OSSOS discoveries 
\citep[see][]{Bannister2016,Lawler2018}.  
The high-quality orbit dynamics and population statistics of OSSOS form a framework on which to overlay the color information provided by Col-OSSOS. The Col-OSSOS sample consists of the 96 TNOs  brighter than or equal to $m_r$ = 23.6 in the 13AE, 13AO, 13BL, 14BH, and 15BS OSSOS blocks.
Col-OSSOS photometry for the 35 TNOs from 13BL and 14BH blocks are reported in this paper (see Section \ref{sec:sample}).


\subsection{Filter Selection and Color Precision}
\label{sec:filters}

The compositional classes described in Section \ref{sec:motivations} are identified by clusters in the optical and NIR reflectance colors exhibited by TNOs. 
The predominantly linear spectra of small, spectrally featureless TNOs in optical wavelengths are fully characterized by their $g-r$ color (see \citealt{Doressoundiram2007}). 
The neutral/red bifurcation of the dynamically excited TNOs exists only in the optical, and is most prominent in $(g-r)$ \citep{Doressoundiram2008,Fraser2012}. 
With observations at wavelengths longer than $I$, additional surface classes become apparent in other small-body populations \citep{DeMeo2009, Emery2011}.
This is shown indirectly for TNOs by \citet{Doressoundiram2008}, who present $BVRI$ photometry of $\sim$100 sources, with only two surface types apparent. 
The transition from the optical to near-infrared spectral gradients occurs at $\sim$1 $\mu$m; thus, only filters at longer wavelengths can provide the necessary slope information. 
It should be noted that \citet{DalleOre2013} present no practical difference between the $J$, $H$, and $K$ bands in terms of identifying their proposed compositional classifications. 

\begin{figure}[htbp]
\includegraphics[width=\textwidth]{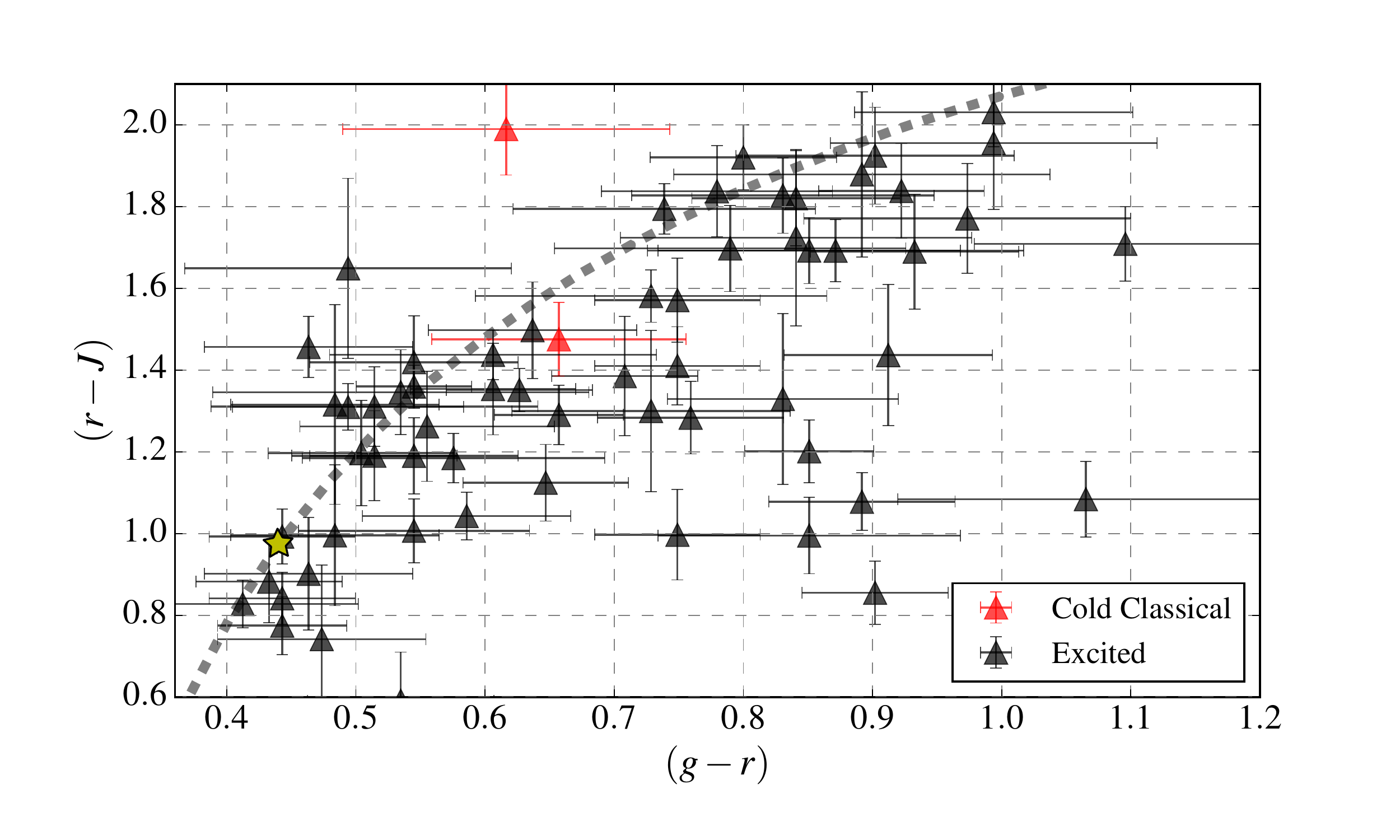}
\caption{\label{fig:pastdata} Non-simultaneous mean observed (converted) $grJ$ color-space measurements of trans-Neptunian populations existing at the conception of the Col-OSSOS survey. Plotted are derived colors for  sources with optical and J-band photometry, as indexed by MBOSS \citep{Hainaut2012} or reported by  \cite{Peixinho2015}. $(g-r)$ was estimated from the reported $(B-V)$ and the conversion of \citet{Jester2005}. $(r-J)$ was estimated by first estimating $(r-i)$ from the reported $(R-I)$ using the conversion reported by \citet{Jester2005} and then estimating $(R-r)$ using the conversion reported by \citet{Jordi2006}. $(R-r)$ is used to estimate $(r-J)$ as $(r-J)$ = $(R-J)$ - $(R-r)$. The Solar color, with g - r = 0.45 and r - J = 0.97, is shown by the yellow star. The dashed curve indicates the reddening line, a  line of constant spectral slope through the $grJ$ spectral range, calculated using the pysynphot software package \citep{Lim+2015}. A broad trend of redness relative to the Sun exists, but detail within the population cannot be discerned.}

\end{figure}

In order to distinguish between the compositional classes proposed by \citet{DalleOre2013} and \citet{Fraser2012}, we selected two optical broad-band filters, $g$ and $r$, and a near-infrared filter, $J$, for our observations. 
These were the fewest filters that could define the optical and near-infrared slope of each TNO in the wavelength region of interest; $(g-r)$ characterizes the optical slope, and $(r-J)$ characterizes the near-infrared slope. Figure \ref{fig:pastdata} shows the color precision of all available  optical+J-band color measurements as of the start of our survey \citep{Hainaut2012,Peixinho2015}, with appropriate conversions to $grJ$ \citep{Jester2005,Jordi2006}. These measurements are mean values, which do not necessarily account for the rotational variability of each TNO. These data demonstrate the importance of precision and temporal near-simultaneity in color measurements: their precision is insufficient to distinguish the color classes discussed in Section \ref{sec:motivations}.
 Those classes have optical and NIR colors that differ by as little as 0.04-0.06 magnitudes from class to class. 
The photometric uncertainty in previously published color studies ranges for 0.04 to 0.2 mag, with no measurements in the size/H range that Col-OSSOS aimed to sample  \citep{Hainaut2012,Peixinho2015}. To achieve 0.06 mag or better photometric  precision for the Col-OSSOS sample required the collecting area of  an 8-10-m class telescope and  the non-standard observing and analysis techniques that we describe in Sections \ref{sec:observing_strategy},  \ref{sec:colours} and Appendixes \ref{sec:optical} and  \ref{sec:NIR}. 

As we observe in a filter system that is close to widely used bandpasses, but has subtle distinctions worth accounting for at our required level of photometric precision, throughout this paper we use the following nomenclature for our filters and corresponding colors:
\begin{description}[labelwidth=1cm, leftmargin=1cm, before={\renewcommand\makelabel[1]{\bfseries ##1}}, itemsep=-1pt]
\item[$g, r, z$] optical bandpasses, in contexts where the specific filter system does not need to be distinguished
\item[$J$]  Maunakea Observatory (MKO) filter set $J$ band \citep{2002PASP..114..169S}, $\lambda$=12500 \AA, 11500-13300 \AA ~coverage
\end{description}
Where necessary for specific observations and color conversions, we specify the exact bandpass in $g$, $r$, or $z$ with appropriate subscripts, as shown here for $r$:
\begin{description}[labelwidth=1cm, leftmargin=1cm, before={\renewcommand\makelabel[1]{\bfseries ##1}}, itemsep=-1pt]
\item[$r_S$] Sloan Digital Sky Survey (SDSS) Photometric System \citep{Fukugita1996, Padmanabhan2008}
\item[$r_G$] Gemini Multi-Object Spectrograph (GMOS) North filters\footnote{See \url{http://www.gemini.edu/node/10420}}. Either acquired with the E2V or the Hamamatsu detectors (specified as needed; see Section \ref{sec:observing_strategy})
\item[$r_{PS}$] Pan-STARRS1 photometric system \citep{Tonry2012}
\end{description}

\subsection{Observing Strategy}
\label{sec:observing_strategy}

Our primary observing facility is the 8.1-m Frederick C. Gillett Gemini North Telescope located on Maunakea, Hawai`i.  
In the optical, each target was observed with the imaging mode ($5.5\arcmin \times 5.5 \arcmin$ FOV) of the Gemini Multi-Object Spectrograph \citep[GMOS;][]{Hook2004} using the r$\_$G0303 ($\lambda$=6300 \AA, $\delta\lambda$=1360 \AA) and g$\_$G0301($\lambda$=4750 \AA, $\delta\lambda$=1540 \AA) filters, similar to the SDSS $r$ and $g$ bands.
The GMOS observations were obtained in 1$\times$1 binning mode. This had $0.0747 \arcsec$ pixels with the e2v deep depletion charge-coupled devices (CCDs) available in GMOS during the 2014-2016 observing semesters, and $0.0807 \arcsec$ pixels after the 2017 installation of Hamamatsu red-sensitive CCDs. GMOS observations were dithered by $3\arcsec$ to $5\arcsec$ from exposure to exposure. For the near-infrared measurements, we observed 
with the Near-Infrared Imager \citep[NIRI;][]{Hodapp2003} in Maunakea Observatory $J$. NIRI observations were acquired using the f/6 camera ($0.116 \arcsec$ per pixel resolution) with a $119.9 \arcsec \times 119.9 \arcsec$ FOV. NIRI observations utilized a grid dither pattern with $8 \arcsec$ spacing between exposures to ensure accurate background measurement and removal.
For both NIRI and GMOS, these instrument configurations allow the best possible characterization of the point spread function (PSF), and hence, the best knowledge of the photometric curve of growth.

For our color precision requirement to detect and distinguish the \citet{DalleOre2013} classes, we aim for an 0.06 mag color precision in $(g-r)$ and $(r-J)$ for all targets. To achieve our desired color precision (Section \ref{sec:filters}), we require a signal-to-noise (SNR) $\geq 25$ in $g$ and $r$, and SNR $\geq 20$ in $J$.  For each Col-OSSOS TNO,  the total effective exposure time required in each filter was estimated  using the mean OSSOS r-band discovery magnitude and assuming very red optical and only modestly red near-infrared colors relative to those typically exhibited by small TNOs (Figure \ref{fig:pastdata}): $g-r =1.1$, and $r-J=1.2$. 
Ideally, most TNOs will have bluer optical colors and redder near-infrared colors, and their observations will thus meet our SNR requirements regardless of their true color.
At the time of the Gemini observations no light curves were yet measured for these TNOs. 
If the 15-20 OSSOS CFHT observations of the TNO over 1-2 years showed significant $\pm 0.3$ mag photometric variability, additional frames in all filters were added to the Gemini observing sequence, in case observing happened during minimum TNO brightness. 

We used Gemini's fast instrument switching abilities ($\lesssim4$~minutes) to provide near-simultaneous colors: all targets were visited in an \emph{unbroken} imaging sequence of $rgJgr$. 
Anchoring the sequence with $r$ exposures lets us account for light curve variations during color estimates (discussed further in Section \ref{sec:colours}). Observations were executed to gather the desired cumulative SNR by taking half the required optical frames  before and  then after the required $J$-band observations. Individual GMOS exposures were set to 300s in duration. 
Individual NIRI exposures were limited to 120s, both to minimize trailing losses and to mitigate the high sky background.

The telescope tracked at the sidereal rate, permitting use of calibration stars within the images (see Appendixes \ref{sec:optical} and \ref{sec:NIR}). Source trailing in each observation was minimal, as Col-OSSOS targets typically have on-sky motions of several arcseconds per hour or less. The timing of the observations was chosen to avoid the TNO passing over or close to bright or contaminating background stars and galaxies. A combination of SDSS observations \citep{York2000}, stacked OSSOS images \citep{Bannister2018},  and stacked Pan-STARRS1 \citep{2016arXiv161205560C,2016arXiv161205243F} images were used to identify times when the TNOs were moving through areas of sky devoid of background sources brighter than $\sim$24th magnitude. The majority of the GMOS observations were made at an airmass $< $2, and most of the NIRI photometry was obtained at an airmass $<$1.4. With Gemini's queue scheduling and priority visitor mode, all observations were made in photometric conditions during dark time, in Gemini's 50 percentile sky background SB 50) criteria. 
The images were executed in Gemini's 70 percentile image quality (IQ 70) or better, achieving image quality typically 0.7$^{\prime\prime}$ or better in the optical and 0.5$^{\prime\prime}$ or better in the NIR for the majority of observations.  Before and after each $rgJgr$ sequence, NIRI photometric calibrator frames were acquired with bright standard stars at different elevations, chosen to encompass the range of elevations spanned during the $J$ observations of each TNO. 
Exposure times for the calibration stars were chosen so as to not saturate the NIRI detector, and exposures were taken in a 9-point dither pattern sequence.  Associated NIRI lamp flats, NIRI dark frames, and GMOS bias observations were also obtained for each night of observing.

\subsubsection{$z$-band Imaging}

A subsample of the Col-OSSOS targets were also imaged  with Gemini in the $z$\_$G0304$ filter (8500-10000 \AA~coverage).
The inclusion of $z$ observations was ad hoc and with no pre-defined minimum SNR requirement.
During exceptional sky conditions (IQ 20, c.f. $\sim 0.4\arcsec$ seeing), we reduced the number of $g$ and $r$ frames, maintaining SNR 25, and added GMOS $z_G$-band observations.
They bracketed the start and end of the optical sequences, with the same 300s exposure time.
We describe the $z_G$ data processing in Appendix \ref{sec:optical} and their implications in \citet{Pike2017}.
Additional simultaneous $z$ images were acquired with Suprime-Cam \citep{Miyazaki2002} on the Subaru Telescope in August 2014. These observations are reported in detail in \citet{Pike2017}.

\section{TNO Color Technique}
\label{sec:colours}

Detailed overviews of the optical and NIR data reduction, photometry, and calibration we performed on Col-OSSOS observations are described in Appendixes \ref{sec:optical} and  \ref{sec:NIR}.   Photometry was performed on each individual GMOS frame. For the NIRI observations, the sequences of images were divided and combined into two stacked images; photometry was performed on each.  Our measurements were made with TRIPPy (Trailed Image Photometry in Python), a dedicated software for photometry of linearly trailed sources \citep{Fraser2016}. TRIPPy makes use of a pill-shaped aperture, an aperture elongated based on a Solar System object's predicted rate of motion. For trailed Solar System sources, this process correctly accounts for the flux that would be lost in making use of circular apertures, while maintaining the photometric precision found with use of small area apertures. For PSFs derived from sidereal tracked stars, aperture corrections can be determined to better than 0.01 mag for the pill aperture  \citep{Fraser2016}. Our optical and NIR measurements were calibrated to SDSS \citep{York2000,Padmanabhan2008} filter system and  Maunakea Observatory (MKO) filter set $J$ band \citep{2002PASP..114..169S} respectively. The estimation of the color transform between the SDSS and Gemini filter sets is described in Appendix \ref{sec:colorterm}. 

The final step to estimating colors is to account for brightness variations of the TNO over the duration of the science sequence. 
Any significant amplitude changes due to rotational variability of the TNO can cause difficulty combining measured broad-band colors if not accounted for. 
Our full GMOS-NIRI-GMOS sequences span between 1 and 6 hours.
Small TNOs ($H > 5$) typically have 6- to 15-hour rotation periods, with peak-to-peak variations of $\sim$0.3 magnitudes \citep{2006AJ....131.1149T,Duffard2009,Benecchi2013}, though significantly larger variations have been observed \citep{Fraser2015,Thirouin2018}, including in the OSSOS sample \citep{Alexandersen2018}.\footnote{\cite{Alexandersen2018} examine the photometric variability for four targets in our first release sample: 2013 UL15, 2013 UP15, 2013 UM15, 2013 UN15.} 

\begin{figure}
\includegraphics[width=\columnwidth]{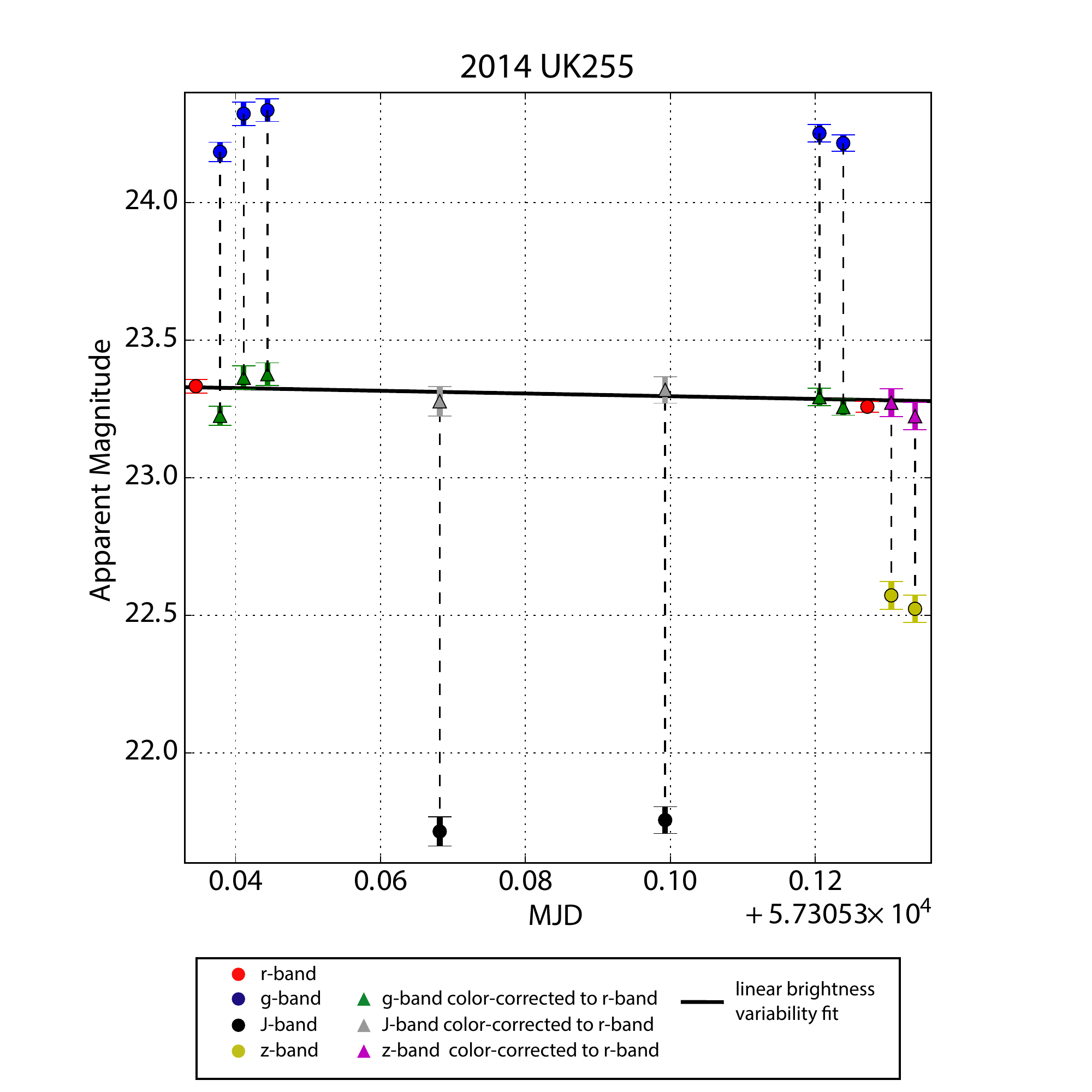}
\caption{\label{fig:light curve}  {Example of observed photometry, and fitted light curve, and colors for object 2014 UK225. Observed $r$, $g$, $z$, and $J$ photometry are shown by circles. Those points adjusted to $r$-band based on the fitted colors are plotted as triangles. The target exhibits an $\sim$0.06 magnitude increase in brightness over the duration of the sequence. This change in brightness is comparable to the precision in measured colors of the source, and is accounted for by our light curve fitting technique.}}
\end{figure}

We took advantage of each sequence's bracketing $g$ and $r$ imagery to correct for light curves by fitting a linear model to all optical data. An example of a fitted light curve for a target that exhibited a variation in brightness over the Col-OSSOS observing sequence is presented in Figure~\ref{fig:light curve}. This model makes two assumptions: that the light curve variations observed across a given sequence are linear in nature, and that the object exhibits no significant color variations across the sequence.  
Thus, the model has as free parameters: slope (change in brightness with time), a reference $r$ magnitude, a color $g-r$, and sometimes a color $r-z$, where $z$ images were available. This model was fit in a least-squares sense to the available data. To evaluate uncertainties on each optical color, we adopt a Monte-Carlo approach. Specifically, each individual photometric measurement was scattered by a Gaussian distribution with width equal to the photometric uncertainty on that point, and the scattered dataset was fit. This process was repeated 200 times, and the the quadrature sum of the standard deviation on the randomized color terms, and the uncertainty on the mean of the individual photometric measurements were adopted as the uncertainty on those points. The fitted linear light curve was then used to estimate the $r$ brightness at the midpoint of each $J$-band measurement so as to determine the mean $r-J$ color. Uncertainties in this color include the photometric uncertainties on $J$, and the uncertainties due to the light curve model parameters (both slope and reference $r$ value).  

We note that the linear fitting process for intrinsic variability will only correctly remove any first derivatives in the light curve. It will not account for the full range of possible photometric variation. We are not able to correct for, nor are we able to meaningfully estimate the additional uncertainty in, our color estimates caused by these unknown non-linear light curve variations. There are undoubtedly shape and albedo effects that impact our color estimates which are not accounted for in the linear fit.  However, With the small number of photometric measurements we have for each target, we have no resolution to further constrain these effects. We discuss the impact of non-linear light curve variations on our results in Section \ref{sec:lc}

\section{Col-OSSOS First Release: The OSSOS 13BL and 14BH Sample}
\label{sec:sample}

The observations we present here comprise a complete flux-limited sample of TNOs: all the $m_r < 23.6$ mag discoveries from the 13BL and 14BH OSSOS survey blocks \citep{Bannister2016,Bannister2018}. In this section, we also present the orbital properties of our TNO release sample and summarize specific observing circumstances and  data analysis unique to these targets compared to the overall Col-OSSOS survey strategy.

\subsection{Observational and Orbital Properties of the TNO Sample from 13BL and 14BH Blocks}

The 35 TNOs  in this first Col-OSSOS release were found in 2013 and 2014 in two OSSOS survey regions of sky near the ecliptic, in a latitude range from the invariant plane up to $5\degr$ off-plane.
They comprise the 18 $m_r < 23.6$ mag discoveries from the 13BL block of OSSOS (\textit{o3l}-designated targets), and the 17 $m_r < 23.6$  mag TNOs from the 14BH block of OSSOS (\textit{o4h}-designated targets). 13BL block is a 20 deg$^2$ region overlaying the invariant plane, centered at R.A. $0^{h}54^{m}$, decl. $+3^{\circ}50$'.
14BH  block is a 21 deg$^2$ region $2-5^{\circ}$ off the invariant plane, centered at R.A. $1^{h}35^{m}$, decl. +$13^{\circ}28$'.
The detection efficiency of OSSOS for moving objects in each survey region is thoroughly characterized \citep{Bannister2018}.

\begin{figure}
\includegraphics[width=0.91\columnwidth]{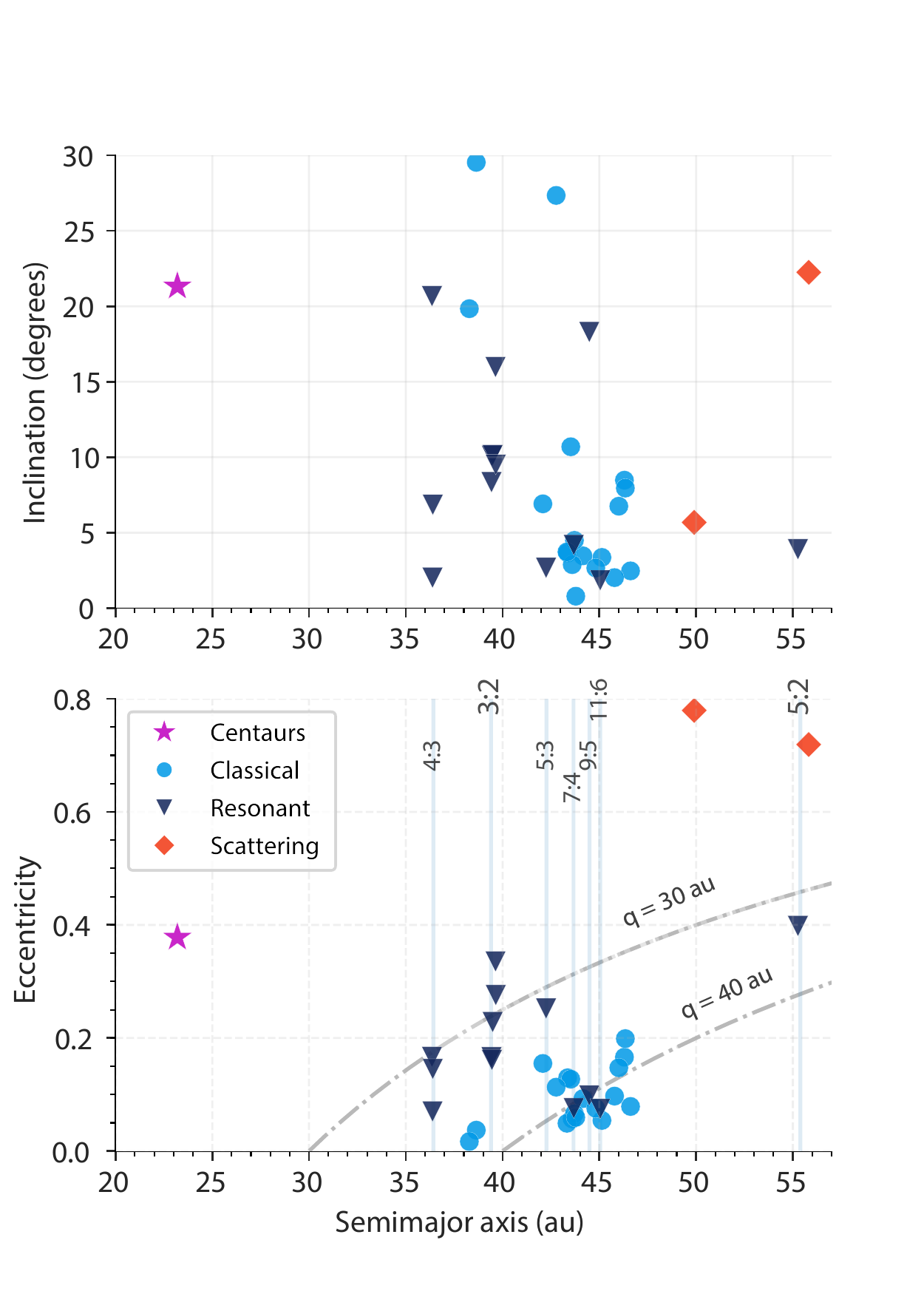}
\caption{\label{fig:sample} Barycentric orbital parameters, derived from \citet{Bannister2018}, of TNOs with Col-OSSOS color measurements presented in this paper. One TNO at $a=130$~au is omitted for better resolution. The 1-$\sigma$ uncertainties are smaller than the size of the plot symbol.}
\end{figure}

Table \ref{tab:sample} lists the orbital information for the 35 TNOs. The orbital distribution is shown in Figure \ref{fig:sample}. We tally the dynamical classes in Table \ref{tab:orbit_numbers}. The TNOs in this first Col-OSSOS release predominantly sample the classical Kuiper belt and the resonant populations. The barycentric orbital properties of the 35 TNOs are derived from 2-5 years of densely sampled observation \citep{Bannister2018}. A full search for binarity of our sample is beyond the scope of this work, but \cite{Fraser2017}, identified three cold classicals in our sample as binaries: 2016 BP81, 2014 UD225, and 2013 SQ99. One hot classical Kuiper belt object, 2013 UQ15, has a semimajor axis, inclination, and eccentricity consistent with being in the dynamical cloud of the Haumea collisional family \citep{Brown2007, Ragozzine2007}.
5 of the 14 resonant TNOs are in the 3:2 mean motion resonance with Neptune, with the rest in the 5:3, 4:3, 7:4, 5:2, 11:6 and 9:5 resonances.
One TNO, 2007 TC434, is securely in the most distant resonance with Neptune yet confirmed, the 9:1 at $a \sim 130$~au \citep{Volk2018}.
There are also 3 objects from the more transient Centaur and scattering populations. The majority of our TNO sample are new discoveries by OSSOS. No targets in our release sample have previous near-simultaneous multi-filter photometry to the precision that we discuss here. Of the few targets with observations earlier than when OSSOS began in 2013, none have published colors \citep[][Kavelaars private communication]{Hainaut2012}. 

All of our 35 targets have absolute magnitudes $H_r > 5$ and are thus much smaller than the dwarf planet size transition of $H\simeq4$ \citep{Brown2008,Tancredi2008}.
Most have $5 < H_r < 8$, brighter than the break in the measured TNO luminosity function at $H_r \sim 8$ \citep{Bernstein2004, Fuentes2008, Fraser2009, Fraser2014,2014AJ....148...55A}.
Only two have $H_r > 10$.

\begin{deluxetable*}{llrrrrrrccclll}
\tabletypesize{\scriptsize}
\tablecaption{\label{tab:sample} Orbital parameters and optical and near-infrared colors of the o3l and o4h Col-OSSOS TNO sample}
\tablehead{
\colhead{MPC} 	& \colhead{OSSOS} 	& \colhead{a} 		&\colhead{e} 	& \colhead{ inc} 		& \colhead{$\Delta$}		& \colhead{r$_{helio}$}	& \colhead{Mean $m_r$} 	& \colhead{$H{_r}$ } 	& \colhead{Orbit} 		& \colhead{$g-r$} 	& \colhead{$r-J$} 	& \colhead{$r-z$} \\
\colhead{ID} 	&  \colhead{ID} 		& \colhead{(au)} 	& \colhead{} 	& \colhead{ ($\degr$)} 	& \colhead{(au)} 	& \colhead{(au)} 	& \colhead{(SDSS)} 			& \colhead{}		& \colhead{Class} 	& \colhead{(SDSS)} 		& \colhead{(SDSS-MKO)} 		& \colhead{(SDSS)} }
\startdata
2013 UR15$^H$  & o3l01 & 55.82 & 0.719 & 22.25 & 15.69 & 16.31 & 23.40 $\pm$ 0.17 & 11.36 & sca & 0.67 $\pm$ 0.02 & 1.49 $\pm$ 0.08 & -- $\pm$ -- \\
2001 QF331$^{F,H}$ & o3l06PD & 42.25 & 0.252 & 2.67 & 31.88 & 32.60 & 22.92 $\pm$ 0.03 & 7.84 & 5:3 & 0.88 $\pm$ 0.03 & 1.67 $\pm$ 0.05 & -- $\pm$ -- \\
2013 US15$^H$ & o3l09 & 36.38 & 0.070 & 2.02 & 33.66 & 34.40 & 23.17 $\pm$ 0.02 & 7.85 & 4:3 & 1.03 $\pm$ 0.02 & 1.49 $\pm$ 0.05 & -- $\pm$ -- \\	
2003 SR317 & o3l13PD & 39.43 & 0.166 & 8.35 & 36.27 & 37.23 & 23.42 $\pm$ 0.08 & 7.77 & 3:2 & 0.64 $\pm$  0.01 & 1.33 $\pm$  0.05 & -- $\pm$ -- \\
2013 SZ99$^H$ & o3l15 & 38.28 & 0.017 & 19.84 & 37.96 & 38.75 & 23.80 $\pm$  0.06 & 7.96 & cla & 0.68 $\pm$  0.02 & 1.35 $\pm$  0.07 & -- $\pm$ -- \\
2010 RE188 & o3l18 & 46.01 & 0.147 & 6.75 & 38.90 & 39.64 & 22.34 $\pm$  0.02 & 6.40 & cla & 0.58 $\pm$  0.02 & 1.43 $\pm$  0.06 & -- $\pm$ -- \\
2013 SP99$^G$ & o3l32 & 43.78 & 0.060 & 0.79 & 41.09 & 42.01 & 23.53  $\pm$  0.05 & 7.35 & cla & 1.00 $\pm$  0.02 & 1.61 $\pm$  0.05 & -- $\pm$ -- \\
2016 BP81$^{B,G,H}$ & o3l39 & 43.68 & 0.076 & 4.18 & 41.81 & 42.54 & 22.83 $\pm$  0.11 & 6.58 & 7:4I & 0.59 $\pm$  0.03 & 1.60 $\pm$  0.07 & -- $\pm$ -- \\
 2013 UL15$^{F,G,H}$ & o3l43 & 45.79 & 0.097 & 2.02 & 42.38 & 43.10 & 23.12 $\pm$  0.10 & 6.82 & cla & 0.90 $\pm$  0.04 & 1.51 $\pm$  0.06 & -- $\pm$ -- \\
2013 UP15$^{G,H}$ & o3l46 & 46.61 & 0.079 & 2.47 & 42.70 & 43.38 & 23.92 $\pm$  0.10 & 7.58 & cla & 0.90 $\pm$ 0.02 & 1.78 $\pm$  0.07 & -- $\pm$ -- \\	
2013 UO15$^G$ & o3l50 & 43.33 & 0.049 & 3.73 & 43.08 & 43.99 & 23.28 $\pm$  0.00 & 6.89 & cla & 0.96 $\pm$  0.02 & 1.75 $\pm$  0.04 & -- $\pm$ -- \\
2013 UM15$^{F,G}$ & o3l57 & 45.04 & 0.075 & 1.84 & 43.77 & 44.46 & 23.39 $\pm$  0.00 & 6.95 & 11:6 & 1.08 $\pm$ 0.01 & 1.56 $\pm$  0.06 & -- $\pm$ -- \\
2006 QF181$^G$ & o3l60 & 44.82 & 0.076 & 2.66 & 44.20 & 44.54 & 23.56 $\pm$ 0.02 & 7.08 & cla & 0.89 $\pm$  0.03 & 1.65 $\pm$  0.05 & -- $\pm$ -- \\
 2013 UN15$^{D,G,H}$ & o3l63 & 45.13 & 0.054 & 3.36 & 44.39 & 45.14 & 24.13 $\pm$ 0.09 & 7.62 & cla & 1.04 $\pm$ 0.03 & 1.57 $\pm$ 0.08 & 0.37 $\pm$ 0.09 \\
2013 UN15$^{C,E,G,H}$ & o3l63 & 45.13 & 0.054 & 3.36 & 44.20 & 45.19 & 23.62 $\pm$ 0.00 & 7.11 & cla & 1.08 $\pm$  0.03 &1.82 $\pm$ 0.04 & 0.73 $\pm$ 0.06 \\
2013 UX18 & o3l69 & 43.60 & 0.057 & 2.89 & 44.78 & 45.73 & 23.93 $\pm$ 0.00 & 7.37 & cla & 0.93 $\pm$ 0.01 & 1.64 $\pm$ 0.08 & -- $\pm$ -- \\
2013 SQ99$^{B,G,H}$ & o3l76 & 44.15 & 0.093 & 3.47 & 46.60 & 47.34 & 23.17 $\pm$ 0.04 & 6.45 & cla & 0.98 $\pm$ 0.03 & 1.51 $\pm$ 0.06 & 0.56 $\pm$ 0.03 \\
2013 UQ15$^{A,F,H}$ & o3l77 & 42.77 & 0.113 & 27.34 & 46.84 & 47.54 & 23.02 $\pm$ 0.22 & 6.28 & cla & 0.53 $\pm$ 0.04 & 0.92 $\pm$ 0.11 & -- $\pm$ -- \\
2013 SA100$^{D,H}$ & o3l79 & 46.30 & 0.166 & 8.48 & 49.69 & 50.43 & 23.02 $\pm$ 0.03 & 6.03 & cla & 0.63 $\pm$ 0.02 & 1.66 $\pm$ 0.08 & 0.44 $\pm$ 0.02 \\
2013 SA100$^{E,H}$ & o3l79 & 46.30 & 0.166 & 8.48 & 49.30 & 50.29 & 22.78  $\pm$ 0.03 & 5.81 & cla & 0.67  $\pm$ 0.02 & 1.48  $\pm$  0.04 & 0.42  $\pm$  0.02 \\
2014 UJ225$^H$ & o4h01 & 23.20 & 0.378 & 21.32 & 17.26 & 17.83 & 23.05 $\pm$ 0.08 & 10.61 & cen & 0.65 $\pm$ 0.01 & 1.08 $\pm$ 0.09 & -- $\pm$ -- \\
2014 UQ229 & o4h03 & 49.90 & 0.779 & 5.68 & 20.84 & 21.83 & 22.80 $\pm$ 0.13 & 9.51 & sca & 1.03 $\pm$ 0.02 & 1.94 $\pm$ 0.05 & -- $\pm$ -- \\
2014 UX229 & o4h05 & 39.63 & 0.335 & 15.97 & 25.76 & 26.41 & 22.50 $\pm$ 0.07 & 8.34 & 3:2 & 0.64 $\pm$ 0.02 & 1.53 $\pm$ 0.06 & -- $\pm$ -- \\
2010 TJ182 & o4h07 & 39.65 & 0.276 & 9.50 & 27.88 & 28.86 & 22.55 $\pm$ 0.02 & 8.02 & 3:2 & 0.60 $\pm$ 0.02 & 1.37 $\pm$ 0.04 & -- $\pm$ -- \\
2014 UV228 & o4h09 & 39.49 & 0.228 & 10.13 & 30.78 & 31.75 & 23.57 $\pm$ 0.05 & 8.62 & 3:2 & 0.65  $\pm$ 0.02 & 1.51 $\pm$ 0.04 & -- $\pm$ -- \\
2014 UO229  & o4h11 & 39.45 & 0.161 & 10.09 & 33.77 & 34.03 & 23.82 $\pm$ 0.03 & 8.52 & 3:2 & 0.72 $\pm$ 0.02 & 1.15 $\pm$ 0.06 & -- $\pm$ -- \\
2014 UD229 & o4h13 & 36.39 & 0.145 & 6.85 & 33.58 & 34.31 & 23.66 $\pm$ 0.02 & 8.35 & 4:3 & 0.71 $\pm$ 0.02 & 1.23 $\pm$ 0.06 & -- $\pm$ -- \\
2014 US229 & o4h14 & 55.26 & 0.398 & 3.90 & 32.32 & 33.31 & 23.47 $\pm$  0.01 & 8.31 & 5:2 & 0.63 $\pm$ 0.02 & 1.50 $\pm$ 0.05 & -- $\pm$ -- \\
2014 UX228 & o4h18 & 36.35 & 0.167 & 20.66 & 37.01 & 37.99 & 23.20 $\pm$ 0.03 & 7.46 & 4:3 & 0.56 $\pm$ 0.03 & 1.44 $\pm$ 0.05 & -- $\pm$ -- \\
2014 UK225$^H$ & o4h19 & 43.52 & 0.127 & 10.69 & 37.09 & 38.06 & 23.32 $\pm$ 0.04 & 7.57 & cla & 0.95 $\pm$ 0.02 & 1.60 $\pm$ 0.04 & 0.69 $\pm$ 0.02 \\
2014 UL225$^H$ & o4h20 & 46.34 & 0.199 & 7.95 & 37.36 & 37.96 & 23.33 $\pm$ 0.11 & 7.57 & cla & 0.55 $\pm$ 0.03 & 1.03 $\pm$ 0.09 & -- $\pm$ -- \\
2014 UH225$^H$ & o4h29 & 38.64 & 0.037 & 29.53 & 39.08 & 40.06 & 23.48 $\pm$ 0.11 & 7.50 & cla & 0.57 $\pm$ 0.02 & 1.69 $\pm$ 0.04 & 0.38 $\pm$ 0.03 \\
2014 UM225$^H$ & o4h31 & 44.48 & 0.098 & 18.30 & 39.50 & 40.16 & 23.53 $\pm$ 0.06 & 7.52 & 9:5 & 0.80 $\pm$ 0.02 & 1.60 $\pm$ 0.04 & -- $\pm$ -- \\
2007 TC434 & o4h39 & 129.92 & 0.695 & 26.47 & 39.74 & 40.60 & 23.47 $\pm$ 0.05 & 7.43 & 9:1 & 0.64 $\pm$ 0.02 & 1.59 $\pm$ 0.05 & -- $\pm$ -- \\
2014 UD225$^{B,G,H}$   & o4h45 & 43.36 & 0.130 & 3.66 & 43.67 & 44.29 & 22.98 $\pm$ 0.05 & 6.55 & cla & 0.74 $\pm$ 0.02 & 1.42 $\pm$ 0.06 & -- $\pm$ -- \\
2001 RY143 & o4h48 & 42.08 & 0.155 & 6.91 & 46.34 & 47.32 & 23.66 $\pm$ 0.08 & 6.95 & cla & 0.92 $\pm$ 0.03 & 1.88 $\pm$ 0.06 & -- $\pm$ -- \\
2014 UE225$^{G,H}$ & o4h50 & 43.71 & 0.066 & 4.49 & 45.96 & 46.56 & 22.92 $\pm$ 0.01 & 6.27 & cla & 1.03 $\pm$ 0.02 & 1.87 $\pm$ 0.05 & -- $\pm$ -- \\
\enddata
\tablenotetext{^A}{Orbit consistent with the Haumea collisional family cluster.}
\tablenotetext{^B}{Confirmed as a binary in \citet{Fraser2017}.}
\tablenotetext{^C}{Photometry includes CFHT measurements.}
\tablenotetext{^D}{2014B observations}
\tablenotetext{^E}{2015B observations}
\tablenotetext{^F}{Telescope tracked non-sidereally at the TNO's rate of motion on-sky.}
\tablenotetext{^G}{Targets with previously-published  optical spectral slopes in \cite{Fraser2017} reprocessed here using the latest version of TRIPPy
\citep{Fraser2016}, the most recent data analysis pipeline, and improved SDSS color terms.}
\tablenotetext{^H}{Targets with previously-published optical colors in \cite{Pike2017}  reprocessed here using the latest version of TRIPPy
\citep{Fraser2016}, the most recent data analysis pipeline, and improved SDSS color terms.  Additional Subaru $r$ and $z$ photometry for some of the highlighted targets is reported in \cite{Pike2017}. Only $(r-z)$ colors obtained from Gemini observing sequences if available are reported here. }
\tablenotetext{}{Orbit Class: Dynamical classification of barycentric orbits from a 10 Myr integration: cen=centaur, sca = scattering disk, cla = classical belt, $N$:$M$= mean motion resonance with Neptune. An `I' after the resonant identifier signifies an insecure resonance classification. See \citet{Bannister2018} for further details.}
\tablenotetext{}{Geometric parameters  and derived $H_r$ are reported for the time of the Col-OSSOS observation (see Table \ref{tab:sample_obs}).}
\tablenotetext{}{Mean $m_r$ is the mean of the measured Col-OSSOS $r$-band photometry. Values for individual frames are reported in Table \ref{tab:sample_obs}.}
\tablenotetext{}{A machine-readable version of this table can be found in the online supplemental files. Those targets with two observation epochs have a separate entry for the color measurements derived at each epoch, ordered chronologically.}
\end{deluxetable*}

\begin{deluxetable}{lll}[h]
\tablecolumns{3}
\tablecaption{\label{tab:orbit_numbers} Dynamical classifications of the Col-OSSOS 13BL and 14BH block targets}
\tablehead{\colhead{Orbital Class} & \colhead{$\#$} & \colhead{Comment} }
\startdata
Centaurs & 1 & \\
Cold Classicals & 10 & $i<5\degr$ inclination, all in main belt \\
Hot Classicals & 8 & 1 consistent with the Haumea family \\
Resonant & 14 & 3:2,  5:3, 4:3, 7:4$^*$, 5:2, 11:6, 9:5, 9:1 \\
Scattering & 2 & \\
\enddata
\tablenotetext{^*}{7:4 resonance identification of 2016 BP81 is insecure. See \citet{Bannister2018} for further details.}
\end{deluxetable}

\subsection{Observations and Data Analysis}
\label{sec:uniqobs}
The Gemini observations were acquired as described in Section~\ref{sec:observing_strategy}, during 2014--2015 (under programs GN-2014B-LP-1 and GN-2015B-LP-1) with some minor exceptions described below.  We also note here any details where the specific analysis differed from the general analysis for the whole survey given in Section \ref{sec:colours} and Appendixes~\ref{sec:optical} and~\ref{sec:NIR}. Table \ref{tab:sample_obs} gives a detailed summary of the GMOS-N and NIRI observations for each target TNO. For all observations, GMOS-N was equipped with the e2v deep depletion detectors. Table \ref{tab:sample_obs} lists the calculated SDSS magnitude of the observed TNO and associated Gemini-filter zero point for each GMOS and NIRI exposure, as detailed in Appendix \ref{sec:colorterm}. We note that our reported uncertainty in the optical magnitudes combines the uncertainties in the SDSS transformation, calculated zero point, and flux measurement. The full photometry sequences for all target TNOs in this paper are presented in the supplemental material, and a representative sample are plotted in Appendix \ref{sec:aphotplots}.

Exceptional sky conditions occurred during the 2014 August priority visitor run, where many observations were acquired in 0.4-0.5\arcsec seeing. The 2015 October priority visitor run had IQ 20 (0.3--0.6$\arcsec$ seeing). When observing in IQ 20 conditions, we shortened the individual GMOS frames to be less than 300s for 10 TNOs (2001 RY143, 2010 RE188, 2010 TJ182, 2013 UR15, 2014 UE225, 2014 UH225, 2014 UK225, 2014 UL225, 2014 UV228, and 2016 BP81) while preserving our desired SNR goals.

Two TNOs in the sample (2013 SA100 and 2013 UN15)  were observed twice. We report both color measurements  in Tables \ref{tab:sample} and \ref{tab:sample_obs}.
The dynamically excited TNO 2013 SA100 was observed in both the 2014B and 2015B semesters, with a full $grzJ$ sequence at each epoch. 
The cold classical 2013 UN15 was observed in both the 2014B and 2015B semesters. As it appeared very red in the 2014B observations ($r-J \sim 1.5$; Table \ref{tab:sample}), additional time was spent on the 2015 $J$ observations, with no Gemini $g$ observations. 
The $g$ observations were instead acquired simultaneously by MegaCam on CFHT with an $rgr$ sequence of ten 300 s g.MP9402 filter images bracketed by two 300 s r.MP9602 filter images before and after. We also report the CFHT photometry and calculated zero points in Table \ref{tab:sample_obs}.

 For 4 of the 35 objects (2001 QF331, 2013 UL15, 2013 UM15, and 2013 UQ15), the telescope instead tracked at the on-sky rate of motion for the target TNO rather than standard sidereal tracking. Given their slow rate of motion of a few arcseconds per hour, the PSF of both the stars and TNO were still quite round in the observations.
Thus, the photometry was measured with the same procedure as the sidereally tracked targets, but the uncertainty in the aperture correction was doubled to 0.02 magnitudes to reflect the small errors induced by the non-sidereal tracking. Additionally, \citet{Fraser2017} identified three of the cold classicals in our sample as binaries: 2016 BP81, 2013 SQ99, and 2014 UD225.
In our GMOS observations, the objects were elongated, but the components were not fully separated.
Photometric apertures with radii of 2.5$\times$FWHM and appropriate aperture corrections were used to ensure the flux of both sources was included in the aperture, for a combined photometric measurement (Table \ref{tab:sample_obs}).

For 2013 UX18, the standard $rgJgr$ sequence is incomplete: the first r$_G$-band observation fell on a faint star and was rejected.   For the 2015B observations of 2013 UN15, the CFHT data were reduced with the OSSOS procedures described in \citet{Bannister2018}.
Photometry was measured using TRIPPy, in the same fashion as for the Gemini data (see Appendix~\ref{sec:optical}). The $g-r$ color was extracted from the CFHT photometry using the line fitting technique described in Section~\ref{sec:colours}, and converted to the SDSS system using conversions provided as part of the MegaPipe pipeline \citep{2008PASP..120..212G}. 
Separately, the line fitting technique was applied to the Gemini photometry to extract $r-z$ and $r-J$ colors in the Gemini system. Finally, the colors were converted to the SDSS system using the $g-r$ color found from the CFHT observations.

\subsubsection{Overlap with previous Col-OSSOS Publications}

Optical colors and optical slopes derived from preliminary analysis of Col-OSSOS observations have been previously published in \cite{Fraser2017} and  \cite{Pike2017} for 22 TNOS in our release sample. These targets are identified in Table \ref{tab:sample}. The same Gemini observations used in those publications are analyzed in this Paper. We only report $(r-z)$ colors that were obtained during the Gemini sequences; the full $(r-z)$ sample including near-simultaneous supplemental Subaru $z$ observations can be found in  \cite{Pike2017}. We were able to acquire Gemini $z$ observations for five targets: 2013 SA100, 2013 SQ99, 2013 UN15, 2014 UH225, and 2014 UK225.  We report the photometry and $(r-z)$ colors for those targets in Table \ref{tab:sample_obs}. The $(r-J)$ color values of our release sample have not been previously published, but we note that preliminary $(g-r)$ and $(J-r)$ colors for the 9 TNOs that overlap with the \cite{Pike2017} sample were plotted in Figure 3 of \cite{Bannister2017} to compare to the near simultaneous $g,r,$ and $J$ photometry obtained for interstellar object 'Oumuamua.  \cite{Marsset2018} utilizes Col-OSSOS  optical colors of different OSSOS blocks that are not part of this first full data release.

The colors and photometry reported here were reprocessed using the latest version of TRIPPy \citep{Fraser2016}, the most recent data analysis pipeline. We note that small differences in the optical colors and slopes of targets reported in  \cite{Fraser2017}, \cite{Pike2017}, and this work are due to updates to TRIPPY,  improved light curve fitting, and improved estimates  of the color transformations from the Gemini filters to the SDSS photometric system, with the inclusion of additional GMOS observations.  Also, during the image reductions of the 2015B Gemini observations of 2013 UN15 that were published in \cite{Pike2017}, the source was contaminated by a background source in the second half of the GMOS sequence. Upon re-reduction for this work, the last image was found to be useable with a sufficiently small $r=0.8\arcsec$ aperture, which avoided the background star. This has caused a small adjustment in the color, and a large improvement in overall color accuracy, which is reflected in the values reported in Tables \ref{tab:sample} and \ref{tab:sample_obs}.

\begin{deluxetable*}{llllllllll}
\tabletypesize{\scriptsize}
\tablecaption{\label{tab:sample_obs}  o3l and o4h Col-OSSOS TNO Sample Observations}
\tablehead{\colhead{MPC} & \colhead{OSSOS} & \colhead{Header} &  \colhead{Reduced} &  \colhead{Filter} &\colhead{MJD} & \colhead{Gemini mag} &  \colhead{zero point} & \colhead{SDSS mag} & \colhead{Exposure$^*$}\\ \colhead{ID} & \colhead{ID} & \colhead{ID} & \colhead{Filename}  &   & & &  & & \colhead{(s)}}
\startdata
2013 UR15 & o3l01 & O13BL3RQ & N20140825S0315.fits & r$\_$G0303 & 56894.43641 & 23.234 $\pm$ 0.022 & 28.251 $\pm$ 0.005 & 23.275 $\pm $0.022 & 300 \\
2013 UR15 & o3l01 & O13BL3RQ & N20140825S0316.fits & g$\_$G0301 & 56894.44091 & 23.937 $\pm$ 0.033 & 28.136 $\pm$ 0.005 & 24.03 $\pm $ 0.033 & 300 \\
2013 UR15 &o3l01 & O13BL3RQ & O13BL3RQ$\_$0.fits & J & 56894.46418 & 21.953 $\pm$ 0.117& 23.905 $\pm$ 0.02 &  -- $\pm$ -- & 1200 \\
2013 UR15 & o3l01 & O13BL3RQ & O13BL3RQ$\_$1.fits & J &56894.47983 & 21.912 $\pm$ 0.094 & 23.907 $\pm$ 0.02 & -- $\pm$ -- & 1320 \\
2013 UR15 & o3l01 &O13BL3RQ & N20140825S0341.fits & g$\_$G0301 & 56894.49351 & 24.12 $\pm$ 0.035 & 28.175 $\pm$0.005 & 24.213  $\pm$ 0.036 & 225 \\
2013 UR15 & o3l01 & O13BL3RQ & N20140825S0342.fits & g$\_$G0301 & 56894.49706 & 23.977 $\pm$0.032 & 28.18 $\pm$ 0.006 & 24.07 $\pm$ 0.033 & 225 \\
2013 UR15 & o3l01 & O13BL3RQ & N20140825S0343.fits & r$\_$G0303 & 56894.50069 & 23.584 $\pm$ 0.029 & 28.271  $\pm$ 0.007 &  23.625  $\pm$ 0.029 & 225 \\
2001 QF331 & o3l06PD & O13BL3SH & N20140823S0289.fits & r$\_$G0303 & 56892.41022 & 22.839 $\pm$ 0.034 & 28.22  $\pm$ 0.005 & 22.892  $\pm$ 0.034 & 300 \\
2001 QF331 & o3l06PD & O13BL3SH & N20140823S0290.fits & g$\_$G0301 & 56892.41473 & 23.702 $\pm$ 0.039 & 28.127  $\pm$ 0.004 & 23.825  $\pm$ 0.039 & 300 \\
2001 QF331 & o3l06PD & O13BL3SH & N20140823S0291.fits & g$\_$G0301 & 56892.41916 & 23.621 $\pm$ 0.038 & 28.133  $\pm$ 0.005 & 23.745  $\pm$ 0.038 & 300 \\
2001 QF331 & o3l06PD & O13BL3SH & O13BL3SH$\_$0.fits & J & 56892.43259 &21.32 $\pm$ 0.063 & 23.852 $\pm$ 0.02 &  --  $\pm$  -- & 840 \\
2001 QF331 & o3l06PD & O13BL3SH & O13BL3SH$\_$1.fits & J & 56892.44375 & 21.197 $\pm$ 0.055 & 23.851 $\pm$ 0.02 &  --  $\pm$  -- & 960 \\
2001 QF331 & o3l06PD & O13BL3SH & N20140823S0308.fits & g$\_$G0301 & 56892.45469 & 23.716 $\pm$0.037 & 28.175  $\pm$ 0.004 & 23.839   $\pm$ 0.037 & 300 \\
2001 QF331 &	 o3l06PD	&  O13BL3SH &	 N20140823S0309.fits  &	 r$\_$G0303 & 56892.45919 & 22.895 $\pm$  0.033 & 28.263 $\pm$  0.004 & 22.948 $\pm$ 0.033  & 	300 \\ 
\enddata
\tablenotetext{}{A machine-readable version of this table in its entirety can be found in the online supplemental files. A portion is reproduced here for guidance regarding its form and content.}
\tablenotetext{}{All raw Gemini data files and calibration files associated with these observations are available via the Gemini Observatory Archive (\url{https://archive.gemini.edu}). Gemini program IDs for these observations are GN-2014B-LP-1 and GN-2015B-LP-1. }
\tablenotetext{}{All raw CFHT data files and calibration files associated with these observations are available via the CFHT Science Archive  (\url{http://www.cadc-ccda.hia-iha.nrc-cnrc.gc.ca/en/cfht/}). CFHT program IDs for these observations are 14BP05 and  15BP05. }
\tablenotetext{}{The Header ID column reflects the OBJECT keyword in the Gemini/CFHT raw FITS headers, as the internal survey designation for target TNOs on occasion may have changed over the course of the OSSOS survey.}
\tablenotetext{}{The reduced FITS files produced in this analysis are available for download at the Canadian Astronomy Data Centre (\url{http://apps.canfar.net/storage/list/ColOSSOS}). The  online repository will be available at the time the manuscript is published.}
\tablenotetext{^*}{For $J$-band, the reported exposure time is the total effective exposure time of the stacked image}
\end{deluxetable*}

\subsubsection{Light Curve Effects}
\label{sec:lc}
The range of brightness variations we observed across a Col-OSSOS sequence (between approximately 1 and 6 hours duration) was 0 to 0.5 mag. The photometric variability of our first release sample is presented in Appendix \ref{sec:aphotplots} and the supplemental material.  As described in Section \ref{sec:colours}, a linear fit was used to remove light curve effects from our color estimates. We checked for violations of our linear model assumptions that may impact the color measurements presented here. In all cases linearity was sufficient to describe the variability we observed. By comparing the color inferred from the first half and last half of the full $rgJgr$ sequences, we found that no objects exhibited detectable spectral differences over the span of our observations. That is, the $(r-J)$ and $(g-r)$ colors that were inferred from the first half, and last half of each sequence were consistent at better than 2$-\sigma$ in all cases. There still may be other brightness variations present (sinusoidal being one possibility of many) that we cannot determine from our photometry and have not been accounted for in our analysis. Such variations within our observations is unlikely to be correlated with the surface properties of the object in such a way such that it then creates, artificially, the correlations between orbit and surface properties that we report in Section \ref{sec:results}.

\section{Results and Discussion}
\label{sec:results}

The $g-r$ and $r-J$ colors of the 35 TNOs in our sample are presented in Figure~\ref{fig:colours} and are reported in Table \ref{tab:sample}. There are some notable features of the optical-NIR color distribution which we discuss in this section, along with a discussion of clear outliers to the majority of the sample. Additionally, we present an analysis of the intrinsic population of objects that belong to the neutral and red classes of dynamically excited Kuiper belt populations.

\subsection{Colors of the First Release Sample}
\label{sec:first_colours}

\begin{figure*}
\includegraphics[width=\textwidth]{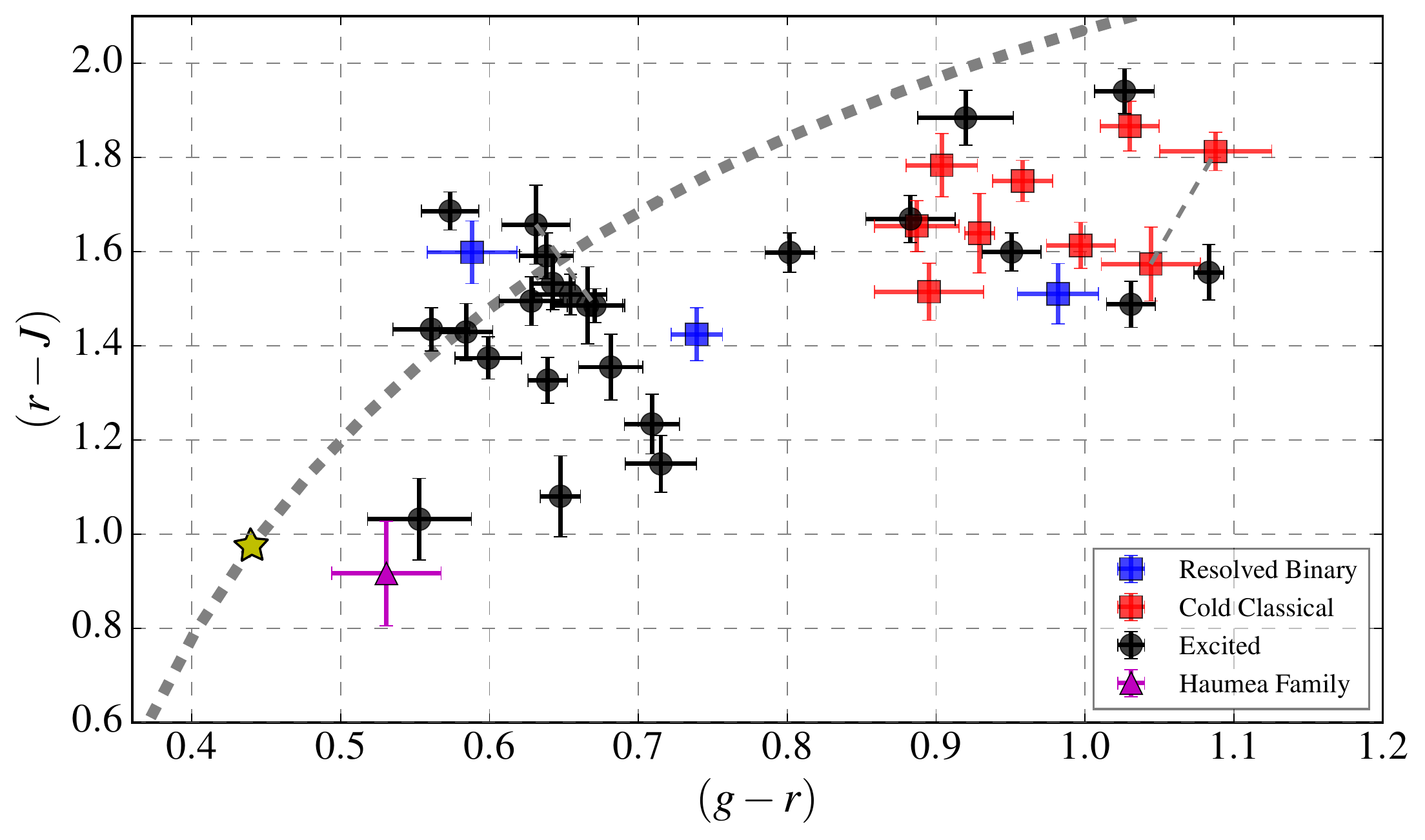}
\caption{\label{fig:colours} Optical and near-infrared colors of the 35 TNOs in the Col-OSSOS first release sample (all $m_r < 23.6$ TNOs in the OSSOS 13BL and 14BH survey blocks). The dashed curve indicates the reddening line; see Section \ref{sec:first_colours} for details. Three objects, 2013 SQ99, 2014 UD225, and  2016 BP81  (blue squares), have cold classical orbits (main Kuiper belt with $i < 5\degr$; red squares) and have been previously identified as widely separated binaries. The object 2013 UQ15 (magenta triangle) is dynamically consistent with the Haumea family.  Excited TNOs (black dots) belong to the centaurs, scattering, resonant, and hot classical dynamical populations. The two measurements of the two re-observed targets, neutral hot classical 2013 SA100 and red cold classical 2013 UN15, are linked by dashed grey lines (the color measurements at each of the two observation epochs are reported in Table \ref{tab:sample}). The Solar color, with $g-r = 0.45$ and $r-J = 0.97$, is shown by the yellow star.}
\end{figure*}

We present the optical and NIR color distribution of Col-OSSOS targets in the 13BL and 14BH OSSOS blocks in Figure~\ref{fig:colours}. 
We include the so-called reddening line, or line of constant spectral slope through the $grJ$ spectral range. This line was calculated using the pysynphot software package \citep{Lim+2015}, using the known bandpass measurements for the SDSS $g$ and $r$ filters, and the Maunakea $J$ filter. We note that the available $J$ bandpass data was measured in the laboratory at room-temperature conditions, rather than at the temperature experienced inside the NIRI dewar\footnote{\url{http://www.gemini.edu/sciops/instruments/niri/imaging/filters}}. This may cause a small deviation of the estimated $J$-band throughput away from the measured curve, and hence the calculated reddening line away from the true curve.

\begin{figure*}
\includegraphics[width=\textwidth]{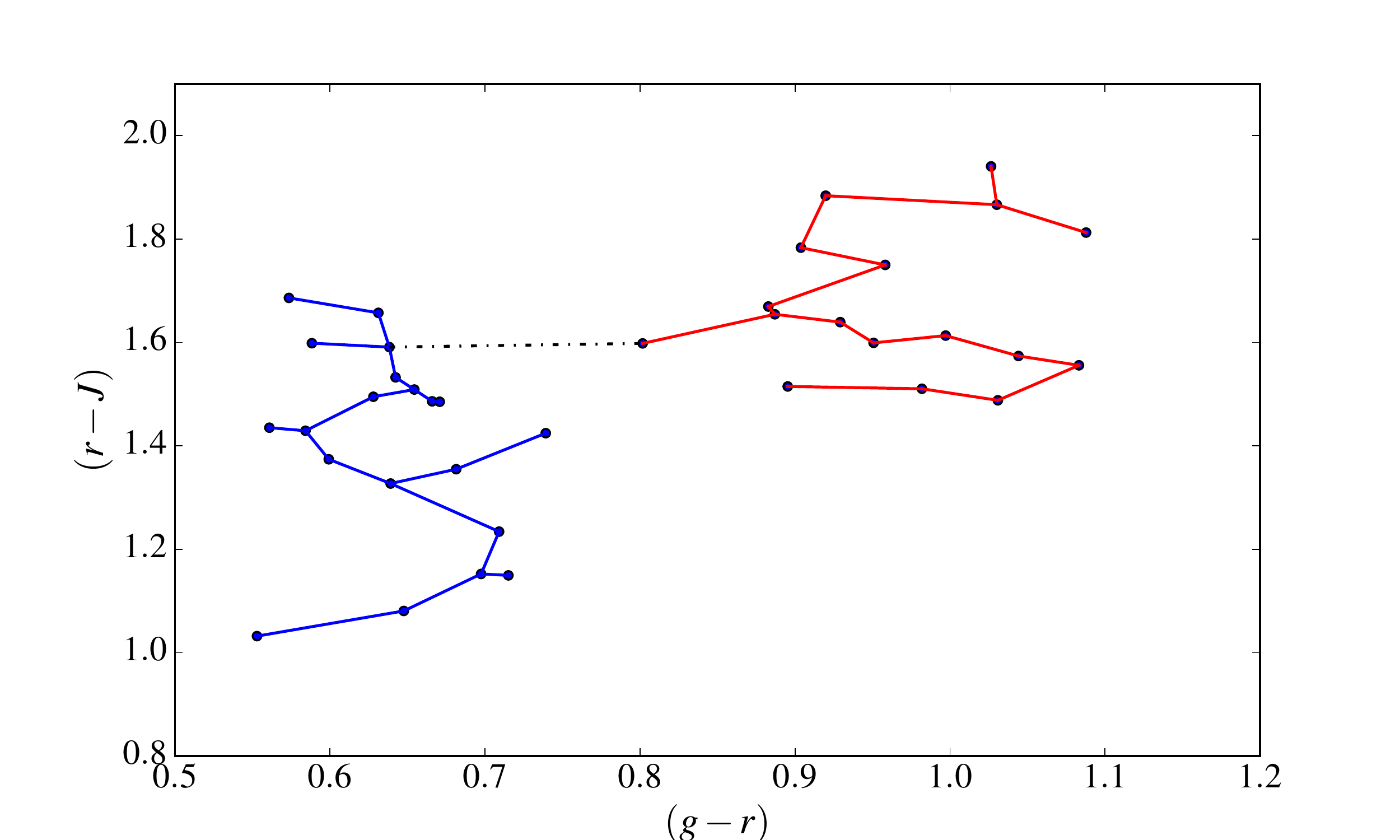}
\caption{\label{fig:FOP} The minimal spanning tree, and sub-trees determined by the FOP-test applied to the optical and NIR colors as shown in Figure~\ref{fig:colours}. The branches connecting members of the two subclasses identified by the test are shown in blue and red lines respectively. The one branch of the full tree that is cut to result in the two sub-trees is shown by the dashed black line.}
\end{figure*}

Sub-structure is apparent in the optical-NIR color space of our 35 TNOs (Figure \ref{fig:colours}). The most notable feature is the bifurcation of the dynamically excited populations into two separate color classes, as seen previously (Section \ref{sec:surfaces}). In our sample, the bifurcation into red and neutral clumps occurs at $g-r \sim0.75$. To test for the presence of the bimodality in the Col-OSSOS observations, we apply a multi-dimensional test, the \emph{F~optimal~plane} (FOP) test developed in \citet{Fraser2012}. This test uses minimal spanning tree clustering in Euclidean color space to test for the significance of potential sub-populations within a dataset. We apply the FOP test to our ($g-r$) and ($r-J$) observations. The FOP test divides the population into two separate classes, that can be approximately divided in just the optical color, that is at ($g-r$)$=0.75$, with only a 2\% chance that such a division would occur by chance. We present the minimum spanning tree generated by the FOP test in Figure \ref{fig:FOP}.

Our sample has two objects that stand as clear outliers from the broad trend of TNO $r-J$ colors being redder than Solar: 2014 UL225 and 2013 UQ15, both of which have $g-r \sim 0.55$ and $r-J \sim 0.95$ (Figure \ref{fig:colours}). 
2013 UQ15 is a hot classical object with orbital elements that place it well within the cloud of objects belonging to the Haumea collisional family \citep{Brown2007,Ragozzine2007}.  The slightly red optical color and neutral NIR color of 2014 UL225 is very similar to those exhibited by known Haumea family members, including 2005 RR43 and 1995 SM55 \citep{Snodgrass2010}, but the deep water ice absorption signature that is characteristic of Haumea family members cannot be identified with the Col-OSSOS observations alone. While 2014 UL225 exhibits similar spectral properties to 2013 UQ15, its orbital inclination of $7.9^{\circ}$ is significantly lower than the Haumea cloud. We consider if resonant diffusion, as experienced by Haumea itself \citep{Ragozzine2007}, could move 2014 UL225 so far away from the orbital phase space occupied by the majority of known family members. Resonant diffusion is most effective in changing the eccentricity of an orbit.  2014 UL225's eccentricity is consistent with having been affected by diffusion, but its low orbital inclination makes resonant diffusion unlikely. \cite{2012Icar..221..106V} examine the long-term orbital evolution of hypothetical Haumea family members, and there are no instances of resonant diffusion down to inclinations less than 20$^\circ$ for stable orbits. 2014 UL225's colors make it an outlier compared to the bulk of the neutral class. Further study is warranted to determine if this object has a water-rich surface and if it is also consistent with the Haumea family.

We find that the optical/NIR colors vary for the two TNOs (2013 SA100 and 2013 UN15) that we observed twice. We report the objects' two sets of colors measurements  in Figure \ref{fig:colours} and Table \ref{tab:sample}. Spectral variability at this level has been observed in other small TNOs \citep[see ][]{Fraser2015}. We note that in neither case do the variations in colors between epochs shift the TNOs from their color class: they are consistent in color class despite the variability. For the hot classical 2013 SA100, the colors measured at each epoch both place the TNO firmly within the neutral class, but the colors are discrepant at the $1$-$\sigma$ level for each of the repeated $g-r$, $r-J$, and $r-z$ colors
Both measurements of cold classical 2013 UN15, place it in the red cloud but the optical and near infrared colors are not consistent within the $1$-$\sigma$ measurement uncertainty. We find 2013 UN15 varies by 0.2 magnitudes or more in  $r-J$ and  $r-z$ \citep[also noted by ][]{Pike2017}.

The bulk of the objects with $g-r <0.75$ (the neutral class) appear to exhibit an inverse correlation in their optical and NIR colors (see Figure \ref{fig:colours}). We apply the Spearman rank test to examine this further.  When we exclude the candidate Haumea collisional fragment 2013 UQ15 as it is a surface type produced via collision and not intrinsic to the planetesimal disk, the Spearman rank test suggests that there is a 30\% chance that the observed correlation would occur by chance, finding no evidence that the correlation is statistically significant. The correlation becomes much stronger if we exclude  2013 UQ15 and 2014 UL225, both outliers from the excited TNO distribution.  Excluding  2013 UQ15 and 2014 UL225, the  Spearman rank test suggests that there is only a 2\% chance that the observed correlation would occur by chance, but it is not clear that  2014 UL225 surface colors are due to some process similar to the Haumea family formation that would justify excluding it from the analysis. Correlations in the optical and NIR colors of dynamically excited TNOs have been identified previously. In particular, the NIR colors of both the neutral and red class members correlate \emph{positively} with their optical colors in the Hubble Space Telescope (HST) Wide-Field Camera 3 broadband filters centered at $\sim$0.6 $\mbox{$\mu$m}$ (F606w), $\sim$0.8$ \mbox{$\mu$m}$ (F814w), and $\sim1.39 \mbox{ $\mu$m}$  (F139m) \citep{Fraser2012}. That result is in stark contrast to the potential inverse optical-NIR color correlation found here in the $g$, $r$, and $J$ filters, with band centers at $0.48 \mbox{ $\mu$m}$, $0.62 \mbox{ $\mu$m}$, and $1.25 \mbox{ $\mu$m}$, respectively.  For the dynamically excited objects in the red class, optical and NIR colors exhibit a positive correlation in the HST filter set. Yet strangely, no correlation of any kind is present in the Col-OSSOS $grJ$ observations of 7 the dynamically excited red objects in our sample (as shown in Figure \ref{fig:colours}). This may suggest that the $J$-band samples a different part of TNO surface reflectance than the F139m HST band, but we cannot demonstrate that the correlation at the level seen in the HST band can be excluded by this sample of TNOs.  A large  $grJ$ sample is required; this topic will be further investigated in future Col-OSSOS data releases.

The color trends observed in the HST filters have been used to provide compositional constraints for the bulk ice and silicate components of small TNOs \citep{Fraser2012}. For example, organic materials have been suggested to account for the positively correlated optical and NIR colors through a simple compositional mix: higher organic content leads to redder optical and NIR colors. If the inverse correlation that we have detected for the neutral class of dynamically excited TNOs is confirmed, then the material that is responsible for the optical-NIR color correlations of the neutral class must exhibit an absorption feature that overlaps $J$, and not the HST F139m filter, so as to account for the signs of the correlations seen in the two different filter sets. Clearly, confirmation of the inverse color correlation in the neutral class is important.

To date, no correlation between the optical and NIR colors of the cold classical objects has been detected. Rather, the bulk of cold classical TNOs exhibits a range of red optical and NIR colors, broadly spanning nearly the full range of colors exhibited by the red dynamically excited objects, though in an uncorrelated fashion. The only objects for which this does not appear to be true are the blue binaries \citep{Fraser2017}.  All members of this recently discovered class of cold classical TNOs exhibit colors compatible with the neutral dynamically excited class throughout the $grzJ$ wavelength range \citep{Pike2017}, and exhibit a nearly 100\% binary fraction \citep{Fraser2017}.  If these objects are indeed survivors of a soft push-out via mean motion resonance sweep-up during the smooth phases of Neptune's outward migration \citep{Fraser2017}, then it follows that the binary cold classical objects  should also exhibit a bimodal optical color distribution, like the dynamically excited TNOs. As yet, insufficient data are available to test this assertion.

\subsection{The Color Fraction of Red/Neutral Surfaces in the Dynamically Excited Kuiper Belt}
\label{sec:colour_fraction}

To estimate the intrinsic fraction of objects in the neutral and red classes of dynamically excited TNOs, we consider an analytic derivation of the number of objects observed within a given OSSOS survey block. To derive that number, we consider a Kuiper belt comprised of bodies that can be described by  object radius $R$ , heliocentric distance $r$, and albedo $a$.  Distributions in those parameters within the Kuiper belt are thus given by $f(R)$, $g(r)$, and $h(a)$, respectively. Here, $g(r)$ will largely depend on the distribution of resonant TNOs, which are preferentially found at certain longitudes with respect to Neptune \citep[see][for example]{Gladman2012}. For resonators and other excited TNOs, no correlations between optical colors and perihelion/argument of perihelion/longitude of ascending node have been observed \citep{Peixinho2015}. Therefore the explicit longitudinal and latitudinal structure will only affect the absolute number of observed objects at a given sky location, and not the fraction of objects in a color class at that pointing. Thus, for clarity of our derivation, we avoid writing $g$ as a function of latitude and longitude. 

The number of objects with albedos between $a$ and $a+da$, radii $R$ and $R+dR$ and distances $r$ and $r+dr$ is given by:

\begin{equation}
n(R,r,a) = A\, f(R)\, g(r)\, h(a)\, da\, dr\, dR
\label{eq:n}
\end{equation}

\noindent
where $A$ is a convenience constant to determine the desired density unit, we adopt objects per square degree.

The magnitude of an object is $m = K - 2.5 \log(a) + 5 \log(r\,\Delta) - 5 \log(R)$ where  $K$ is a constant related to the Solar luminosity, $\Delta$ is the geocentric distance to the object, and is a function of $r$, and we have ignored phase effects. Writing $R$ in terms of $m$ and its derivative with respect to $m$, we have

\begin{eqnarray}
R & = & \frac{r\, \Delta}{\sqrt a}10^{\frac{K-m}{5}} \\
dR & = & \frac{-\ln 10}{5} R dm
\label{eq:R}
\end{eqnarray}

The majority of Col-OSSOS targets have r-band absolute magnitudes brighter than $H_{r}\sim8$, the approximate magnitude at which absolute magnitude distribution of the dynamically excited objects transitions from a steep power-law, to a shallower slope \citep{Bernstein2004, Fuentes2008, Fraser2009, Fraser2014,2014AJ....148...55A}. As such, we will consider only objects with $H_{r}<8$, and approximate the size distribution is a power-law of the form $f(R) = C\, R^{-q}$, where $C$ is a normalization constant and $q$ is the power-law slope. Substituting Equation \ref{eq:R} by this size distribution into Equation \ref{eq:n}, $n$ can be defined as:

\begin{equation}
n(R,r,a) = \frac{\ln10\, C\, A}{5} \, R^{1-q}\, g(r)\, h(a)\, da\, dr\, dm
\end{equation}

Assuming that the Kuiper belt is bounded by distances $r_0\leq r \leq r_1$, and objects in it have albedos with values $a_0 \leq a \leq a_1$, the number of objects between magnitudes $m_0$ and $m_1$ is shown by Equation \ref{eq:16}.

\begin{equation}
N = \frac{ \ln10\, C\, A }{5}\int_{a_o}^{a_1} \int_{r_o}^{r_1} \int_{m_o}^{m_1} h(a) a^{\left(\frac{q-1}{2}\right)}  g(r) r^{1-q} \Delta^{1-q} 10^{\frac{(1-q)(K-m)}{5}} dm\, dr\, da.
\label{eq:16}
\end{equation}

If we substitute $q = 5\alpha+1$, where  $\alpha$ is the logarithmic slope of the power law, we find Equation \ref{eq:17},

\begin{eqnarray}
N & = & \frac{ \ln10\, C\,A }{5}\int_{a_o}^{a_1} \int_{r_o}^{r_1} \int_{m_o}^{m_1} h(a) a^{\frac{5\alpha}{2}}  g(r) r^{-5\alpha} \Delta^{-5\alpha} 10^{\alpha(m-K)} dm\, dr\, da \nonumber \\
& = & \frac{C\,A\, 10^{-\alpha K} }{5\alpha}\int_{a_o}^{a_1} \int_{r_o}^{r_1} h(a) a^{\frac{5\alpha}{2}}  g(r) r^{-5\alpha} \Delta^{-5\alpha} dr da \left[10^{\alpha m_1} - 10^{\alpha m_o}\right]
\label{eq:17}
\end{eqnarray}

We consider a simple survey in which $m_2 >> m_1$ and has a constant efficiency $\eta$ that goes to zero at magnitude $m$. Then the number of observed objects is given by Equation \ref{eq:18}, and we have arrived at the general form of the cumulative luminosity function $N(<m) = 10^{\alpha\left(m-m_o\right)}$, where $\alpha\sim0.7$ \citep{Fraser2008,Fuentes2008,Petit2011}.

\begin{equation}
N (<m) = \frac{C\,A\, 10^{-\alpha K} }{5\alpha}\int_{a_o}^{a_1} h(a) a^{\frac{5\alpha}{2}} da \int_{r_o}^{r_1}  g(r) r^{-5\alpha} \Delta^{-5\alpha} dr 10^{\alpha m}
\label{eq:18}
\end{equation}

Now consider that the Kuiper belt  exhibits two main color populations: the red and neutral objects. These populations differ in their albedo distributions \citep{Stansberry2008,Fraser2014,Lacerda2014}, and overall number density. If, within a given survey pointing, we assume they share the same size and radial distributions, we can derive the observed red:neutral population ratio for a given intrinsic ratio. While the latter assumption hasn't been tested, no detectable size distribution differences, other than absolute number, have been detected over the observable range of the dynamically excited neutral and red classes in surveys that are sensitive to those differences \citep[e.g.][]{Wong2017}. With these assumptions, if the intrinsic ratio of objects in the red and neutral populations are given by $A_{\textrm{n}} = \gamma A_{\textrm{r}}$,  then the \emph{observed} ratio of the red and neutral populations, $R_{\textrm{r,n}}(<m)$), is:

\begin{equation}
R_{\textrm{r,n}}(<m) = \frac{1}{\gamma} \frac{\int_{a_o}^{a_1} h_{\textrm{r}}(a) a^{\frac{5\alpha}{2}} da}{\int_{a_o}^{a_1} h_{\textrm{n}}(a) a^{\frac{5\alpha}{2}} da}
\label{eq:ratio}
\end{equation}

\noindent
For our observations, the limiting magnitude, $m$ is $m_r = 23.6$ though we note that there is no explicit dependence of Equation~\ref{eq:ratio} on $m$.

While it is certainly true that the red and neutral populations exhibit a range of albedos, the true distribution is currently unknown. Thus, for simplicity and a basic first estimate of the intrinsic red:neutral fraction, we model the two populations as having a single unique albedo, $a_{\textrm{r}}$ and $a_\textrm{n}$. That is, $h_\textrm{r}(a) = \delta(a_\textrm{r} - a)$ and $h_\textrm{n}(a) = \delta(a_\textrm{n} - a)$, where $\delta$ is the Dirac delta function. Then we are presented with the simple red:neutral fraction relation

\begin{equation}
R_{\textrm{r,n}}(<m) = \frac{1}{\gamma} \frac{ a_\textrm{r}^{\frac{5\alpha}{2}}}{a_\textrm{n}^{\frac{5\alpha}{2}}}
\label{eq:redneutral}
\end{equation}

\noindent
which, importantly, is independent of limiting magnitude. The mean albedos for the red and neutral populations are $a_\textrm{r}=12\%$ and $a_\textrm{n}=6\%$ respectively \citep{Fraser2014,Lacerda2014}. Thus, we find $R_{r,n} \sim \frac{3.4}{\gamma}$, where $\gamma$ is the \emph{intrinsic} ratio of neutral to red objects in the dynamically hot TNO population. 

As we are considering only the bulk of the dynamically excited TNOs, we exclude the potential Haumea  family member 2013 UQ15 due its unique surface properties attributed to its collisional origin. We avoid counting the three targets (2014 UJ225, 2014 UQ229, and 2013 UR15) which have $H_{r}>8$, and therefore avoid the region where a single power-law size distribution is not satisfied \citep{Bernstein2004, Fuentes2008, Fraser2009, Fraser2014}. Thus, in the remaining bulk sample of dynamically excited objects, there are 9 neutral and 3 red class objects in 14BH block, and 4 neutral and 3 red objects in 13BL block. The red:neutral ratios of each block are consistent at the $2$-$\sigma$ level. Together, the observed ratio is $R_{r,n}=6/13$. Considering the $1$-$\sigma$ 
 range on the observed ratio and Equation~\ref{eq:redneutral}, we find $\gamma=7.4_{-3}^{+3.6}$. Thus, the observed population implies that in the intrinsic population, the neutral class outnumbers the red class, by a factor of  4.4-11.0. We further note that adoption of a distribution of albedos for each class has a tendency to increase this factor substantially. For example, if we adopt uniform albedo distributions that span the observed range of albedos of the neutral and red classes ($0.04\leq a \leq 0.08$ and  $0.08\leq a \leq 0.22$), the inferred intrinsic ratio would be a factor of $\sim3$ higher than what we infer using the mean albedos of each class.  Thus, our result should be interpreted as a lower limit.

\citet{Wong2017} adopt a different approach to determining the relative neutral:red population fraction. We note that the two color categories in \citet{Wong2017}, ``red" and ``very red", are similar to our ``neutral" and ``red" color categories respectively. Instead of integrating to a certain limiting magnitude, \citet{Wong2017} consider only objects detected in their survey to a given size, with appropriate assumptions on albedo of each of the red and neutral classes. Over a similar size range as that discussed here, they find that the intrinsic neutral to red number ratio is $\gamma=3.6\pm1.2$ where the uncertainty on this number is derived from the $1$-$\sigma$ Poisson range on the observed number of objects. This number is in $2$-$\sigma$ agreement with our measured value.

\subsection{The Structure of the Protoplanetesimal Disk}
\label{sec:disk}

If the separate classes of TNOs reflect the compositional structure of the protoplanetesimal disk from which they originated \citep[see][for example]{Fraser2012}, the presence of only two classes of dynamically excited TNO argues for a moderately compositionally homogenous disk, up to the level of our measurement uncertainty.To explain our observed optical-NIR color distribution of excited TNOs, only one compositional division would be needed, between $\sim20$ and $\sim30$~au where the majority of dynamically excited TNOs originated \citep[e.g.][]{2005Natur.435..466G,Levison2008,2013Icar..225...40B,Nesvorny2015b,2016ApJ...825...94N}. We note that we cannot rule out further finer color-composition structure below our measurement precision ($\sim$0.04 mag ) that may exist in the dynamically excited TNO source population. 

From our calculated $\gamma$, we can estimate where the division between the neutral and red classes occurred. To that end, we assume a simple disk, with a surface density described as $\Sigma(r)\propto r^{-\beta}$ with inner and outer extents of $r_{\textrm min}$ and $r_{\textrm max}$. To gauge the radial extent, we turn to models of the Solar System's large scale dynamical restructuring (Section \ref{sec:history}). 
The currently favored scenario for giant planet migration and dispersal of the planetesimal disk is that of \citet{Nesvorny2015}, in which Neptune originates at 22 au, implying a disk inner edge of $r_{\textrm min}\sim23$~au. Objects in the dynamically excited populations originate inside the final location of the 3:2 mean motion resonance with Neptune, or $\sim39$~au. 
We adopt that value for the outer edge of the disk from which dynamically excited objects originated. 
In the disk, we hypothesize a sharp transition distance, $r_{\textrm s}$ for the original locations of the neutral and red populations. We note that our disk model is based on a simple assumption about the radial surface density distribution that at some level is not correct. For example, it is generally accepted that a sharp density gradient at $\sim$30 au is required to halt Neptune during its late stages of migration \citep[see for example][]{2004Icar..170..492G}. Such a gradient is likely steeper than reflected in our simple model. As a result, the distance $r_s$ is likely interior to the value we estimate below.

Additional observational constraints also help inform the portrait of the  protoplanetesimal disk. The dynamically quiescent cold classicals exhibit a different range of $r-z$ colors distinct from red dynamically excited TNOs even though they exhibit similar $r-J$ and $g-r$ colors \citep{Pike2017}. Thus, \cite{Pike2017} infer that the cold classicals are their own unique TNO surface type.
It follows from this result that the disk had a second division beyond which the cold classicals originated.
This second compositional division/boundary must have been near the current inner edge of the cold classical objects, to explain why cold classical-like surfaces are rare (or not present) in the dynamically excited populations. \cite{Fraser2017} found that the blue binary cold classicals, which have neutral colors consistent with the neutral excited TNO surfaces, are interlopers emplaced during Neptune migration. This places an additional constraint on where this cold classical surface boundary can be. Dynamical modeling by \cite{Fraser2017} find that in order to deliver  the blue binaries onto cold classical  orbits  during Neptune migration, neutral surfaces were present up to the inner edge of the cold classical belt. Thus, the transition to red cold classical surfaces would be expected near the start of the present-day cold classical belt with  red excited TNOs  originating more inward than the neutral TNO surfaces. 

Combining our results with these additional observational constraints, we can explore the red/neutral transition region for the source of the excited TNOs. Despite not knowing $\beta$, the power-law slope of the disk surface density, we can use the inferred intrinsic neutral to red population ratio to place some constraints on the transition distance. For $0\leq\beta\leq3$, our values of gamma imply $37.4\leq r_{\textrm s}\leq 38.5$ au if the neutral class originated inside $r_{\textrm s}$, or $32.5\leq r_{\textrm s}\leq 33.4$ au if the neutral class originated outside $r_{\textrm s}$. 
This estimate fails to account for any variation in efficiency of scattering from certain regions of the disk into different dynamical classes within the Kuiper belt. 
It also fails to account for the currently unexplained sharp transition in surface density of the protoplanetesimal disk at $\sim30$~au, which is seemingly required to halt Neptune's migration at the correct distance. As such, this estimate should be taken only as a rough guide for the location of $r_{\textrm s}$. A more thorough estimate will be made through the use of the OSSOS survey simulator \citep{Lawler2018}, and forward-modeling migration model output, when a larger sample of Col-OSSOS photometry is complete.

\begin{figure*}
\includegraphics[width=\textwidth]{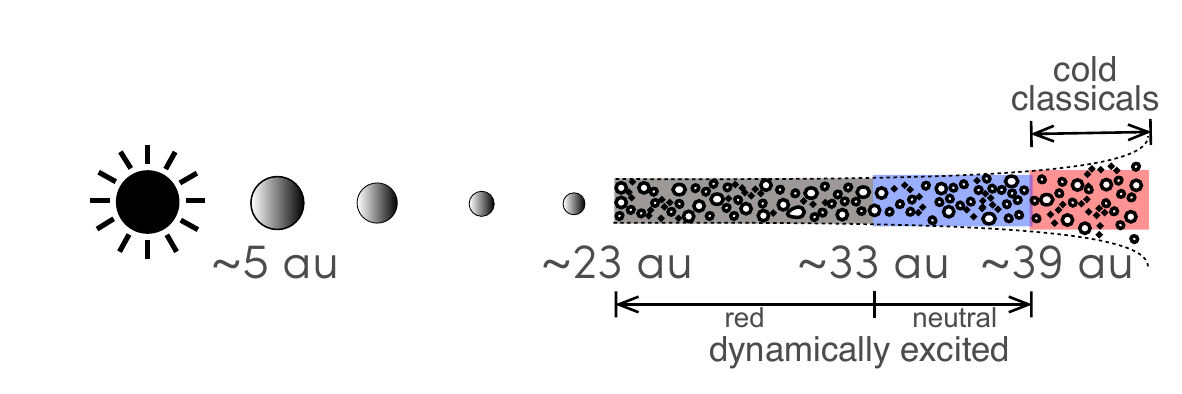}
\caption{\label{fig:disk} The implied planetesimal disk structure, not to scale, under the assumption of a moderately compositionally homogenous disk.}
\end{figure*}

Synthesizing all the observational constraints from \cite{Pike2017} and \cite{Fraser2017}  with our results based on the assumptions described in Section \ref{sec:colour_fraction}, we find a protoplanetesimal disk with a red-blue-red structure, as shown in Figure \ref{fig:disk}.
Closest to the Sun, today's dynamically excited red class originates at a point interior to the neutral class, with the division between the two at $\sim33$~au. 
The dynamically excited neutral class starts interior to the cold classical objects, with a division between the two only a few au inside of the current inner edge of the cold classical region, at $\sim40$~au. 
The higher inclinations of the neutral dynamically excited TNOs \citep{Marsset2018} imply they have experienced a more agitated dynamical history than the red ones \citep{Gomes2003}, potentially complicating this picture. 
A more detailed comparison between Col-OSSOS observations and output of dynamically compatible migration simulations will test the viability of this overall compositional picture. We also note that  finer $grJ$ color-composition structure that is not resolvable by our measurements may exist within each of these composition classes, further complicating this picture.

\section{Conclusions}
We present optical and NIR colors of 35 TNOs, found in the 13BL and 14BH OSSOS discovery blocks with magnitudes brighter than $m_r=23.6$. In $g-r$ and $r-J$, the dynamically excited TNOs, exhibit two classes of objects: the neutral and red classes. We find a tentative negative correlation between the $g-r$ and $r-J$ colors for the neutral class, but additional observations are needed to confirm. 
We find no evidence for a correlation in the colors of the dynamically excited red class. Assuming a  population density that is a separable function of distance, size, and albedo,  we find that the neutral class outnumbers the red class by at least 4.4:1.0 but could be as high as 11.0:1.0.

We find that the cold classical TNOs predominantly occupy the same range of colors in $g-r$ and $r-J$ as the dynamically excited, red class of TNO.  As shown by \citet{Pike2017} however,  the cold classical TNOs occupy a different range of $r-z$, demonstrating that the cold classicals present a different surface than the equivalently optically red excited objects. Combining this observation with our data, we find that within our measurement uncertainty, our observations are consistent with the bulk of TNOs: dynamically excited neutral, dynamically excited red, and cold classical. This excludes rare objects such as the Haumea family members \citep{Brown2007}, the volatile-bearing dwarf planet-sized bodies \citep{Schaller2007,Brown2008}, or the silicate-rich TNO 2004 EW95 \citep{Seccull2018}. 

Based on the assumptions and simple TNO model described in Section \ref{sec:first_colours},  we find our observations are consistent with a planetesimal disk with two compositional divisions separating three separate classes of objects occurred at roughly 33~au, and at just a few astronomical units inside $40$~au, the current inner edge of the cold classical region.  We note  that further finer  color-composition structure in the planetesimal disk in $grJ$ color space that are not resolvable with our measurement cannot be ruled out. To probe the possibility of finer structure in the  TNO color/composition space will require measurement uncertainties smaller than 0.01 magnitude in $g,r,$ and $J$. Our observations and past TNO color measurements are consistent with the 3 surface type model for the bulk of the TNO population. The presence of neutral class interlopers in the cold classical region suggests that the neutral objects bordered the inner primordial edge of the cold classical objects, and the red TNOs began interior to the neutral class. The complete Col-OSSOS sample is expected to include 96 objects from five OSSOS blocks, and will include additional $u$-band photometry from CFHT, acquired simultaneously alongside the Gemini observations. This future four-band dataset will be used to generate a robust taxonomic system for TNOs that accounts for the correlated optical and NIR colors they exhibit.

\section*{Acknowledgements}

The authors acknowledge the sacred nature of Maunakea, and appreciate the opportunity to observe from the mountain.
This work is based on observations from the Large and Long Program GN-2014B-LP-1 and GN-2015B-LP-1, at the Gemini Observatory, which is operated by the Association of Universities for Research in Astronomy, Inc., under a cooperative agreement with the NSF on behalf of the Gemini partnership: the National Science Foundation (United States), the National Research Council (Canada), CONICYT (Chile), Ministerio de Ciencia, Tecnolog\'{i}a e Innovaci\'{o}n Productiva (Argentina), and Minist\'{e}rio da Ci\^{e}ncia, Tecnologia e Inova\c{c}\~{a}o (Brazil).  This work is also based on observations obtained with MegaPrime/MegaCam, a joint project of CFHT and CEA/DAPNIA, at the Canada-France-Hawaii Telescope (CFHT) which is operated by the National Research Council (NRC) of Canada, the Institut National des Sciences de l'Univers of the Centre National de la Recherche Scientifique of France, and the University of Hawaii. We thank the Gemini North staff for their support of the Col-OSSOS program. In particular, we acknowledge the Gemini North queue coordinators, program contact scientists, science operations specialists, and the NIRI and GMOS instrument teams for their assistance. We also thank the CFHT staff for their support of the Col-OSSOS program. The authors also thank the anonymous reviewer for the careful and constructive review that improved this manuscript.

MES was supported by Gemini Observatory and also in part by an Academia Sinica Postdoctoral Fellowship.
M.T.B. appreciates support during Col-OSSOS from UK STFC grants ST/P0003094/1 and ST/L000709/1, the National Research Council of Canada, and the National Science and Engineering Research Council of Canada.  
N.P. acknowledges funding from the Portuguese FCT --- Foundation for Science and Technology (ref: SFRH/BGCT/113686/2015). CITEUC is funded by National Funds through FCT --- Foundation for Science and Technology (project: UID/ Multi/00611/2013) and FEDER - European Regional Development Fund through COMPETE 2020 - Operational Programme Competitiveness and Internationalisation (project: POCI-01-0145-FEDER-006922).
KV acknowledges support from NASA grants NNX15AH59G and NNX14AG93G.

This research used the facilities of the Canadian Astronomy Data Centre operated by the National Research Council of Canada with the support of the Canadian Space Agency.
This work also made use of the Gemini Observatory Archive, NASA's Astrophysics Data System Bibliographic Services, the JPL HORIZONS web interface (\url{https://ssd.jpl.nasa.gov/horizons.cgi}), and data and services provided by the International Astronomical Union's Minor Planet Center.  This research made use of Astropy, a community-developed core Python package for Astronomy \citep{Astropy2013, Astropy2018}. PyRAF are products of the Space Telescope Science Institute, which is operated by AURA for NASA.

The Pan-STARRS1 Surveys (PS1) and the PS1 public science archive have been made possible through contributions by the Institute for Astronomy, the University of Hawaii, the Pan-STARRS Project Office, the Max-Planck Society and its participating institutes, the Max Planck Institute for Astronomy, Heidelberg and the Max Planck Institute for Extraterrestrial Physics, Garching, The Johns Hopkins University, Durham University, the University of Edinburgh, the Queen's University Belfast, the Harvard-Smithsonian Center for Astrophysics, the Las Cumbres Observatory Global Telescope Network Incorporated, the National Central University of Taiwan, the Space Telescope Science Institute, the National Aeronautics and Space Administration under Grant No. NNX08AR22G issued through the Planetary Science Division of the NASA Science Mission Directorate, the National Science Foundation Grant No. AST-1238877, the University of Maryland, Eotvos Lorand University (ELTE), the Los Alamos National Laboratory, and the Gordon and Betty Moore Foundation.

Funding for the Sloan Digital Sky Survey IV has been provided by the Alfred P. Sloan Foundation, the U.S. Department of Energy Office of Science, and the Participating Institutions. SDSS-IV acknowledges support and resources from the Center for High-Performance Computing at the University of Utah. The SDSS web site is \url{www.sdss.org}. SDSS-IV is managed by the Astrophysical Research Consortium for the Participating Institutions of the SDSS Collaboration including the  Brazilian Participation Group, the Carnegie Institution for Science,
Carnegie Mellon University, the Chilean Participation Group, the French Participation Group, Harvard-Smithsonian Center for Astrophysics,
Instituto de Astrof\'isica de Canarias, The Johns Hopkins University,
Kavli Institute for the Physics and Mathematics of the Universe (IPMU) /
University of Tokyo, Lawrence Berkeley National Laboratory,
Leibniz Institut f\"ur Astrophysik Potsdam (AIP),
Max-Planck-Institut f\"ur Astronomie (MPIA Heidelberg),
Max-Planck-Institut f\"ur Astrophysik (MPA Garching),
Max-Planck-Institut f\"ur Extraterrestrische Physik (MPE),
National Astronomical Observatories of China, New Mexico State University,
New York University, University of Notre Dame,
Observat\'ario Nacional / MCTI, The Ohio State University,
Pennsylvania State University, Shanghai Astronomical Observatory,
United Kingdom Participation Group,
Universidad Nacional Aut\'onoma de M\'exico, University of Arizona,
University of Colorado Boulder, University of Oxford, University of Portsmouth,
University of Utah, University of Virginia, University of Washington, University of Wisconsin,
Vanderbilt University, and Yale University.

\appendix

\section{Optical Data Reduction, Photometry, and Calibration}
\label{sec:optical}
In this section, we describe the data processing of the GMOS observations used in our analysis. We also describe the photometry and calibration of our optical measurements from the Gemini filter system to SDSS \citep{York2000,Padmanabhan2008} filter system.

 \subsection{GMOS Data Reduction}

The GMOS images were reduced using the Gemini IRAF package\footnote{\url{http://www.gemini.edu/sciops/data-and-results/processing-software/download}} \citep{2016ascl.soft08006G} and packages from the Ureka Python and PyRAF environment\footnote{\url{http://ssb.stsci.edu/ureka/}}. For each set of observations, we obtained master twilight flats and bias files\footnote{obtained through the Gemini Science Archive (\url{http://www.cadc-ccda.hia-iha.nrc-cnrc.gc.ca/en/gemini/}) and Gemini Observatory Archive (\url{https://archive.gemini.edu/})} in the appropriate binning and bandpass. These master calibration files are produced by Gemini regularly using a large number of individual biases and flats; for each target, we used master calibration files from within a few weeks of the observations. The GMOS observations were adjusted for CCD amplifier gain, bias subtracted and then divided by the master twilight flat field. This removed the vast majority of the structure in the image. \citet{Boone2018} found that the readout of e2v CCDs  may create differential offsets up to 4.5 ADU in the background count when comparing empty pixels of the CCD image to those with flux sources. 
The sky background of our images was much higher than 4.5 ADU, thus the impact of this effect should be negligible on Col-OSSOS photometry derived from GMOS-N e2v imagery.
Each image was reviewed by eye, and those rare frames where the TNO's PSF was blended with a faint background galaxy or star were rejected from our later analysis.

\subsection{Optical Photometry}
\label{sec:optical_phot}
\begin{figure*}
\begin{center}
\includegraphics[width=0.9\textwidth]{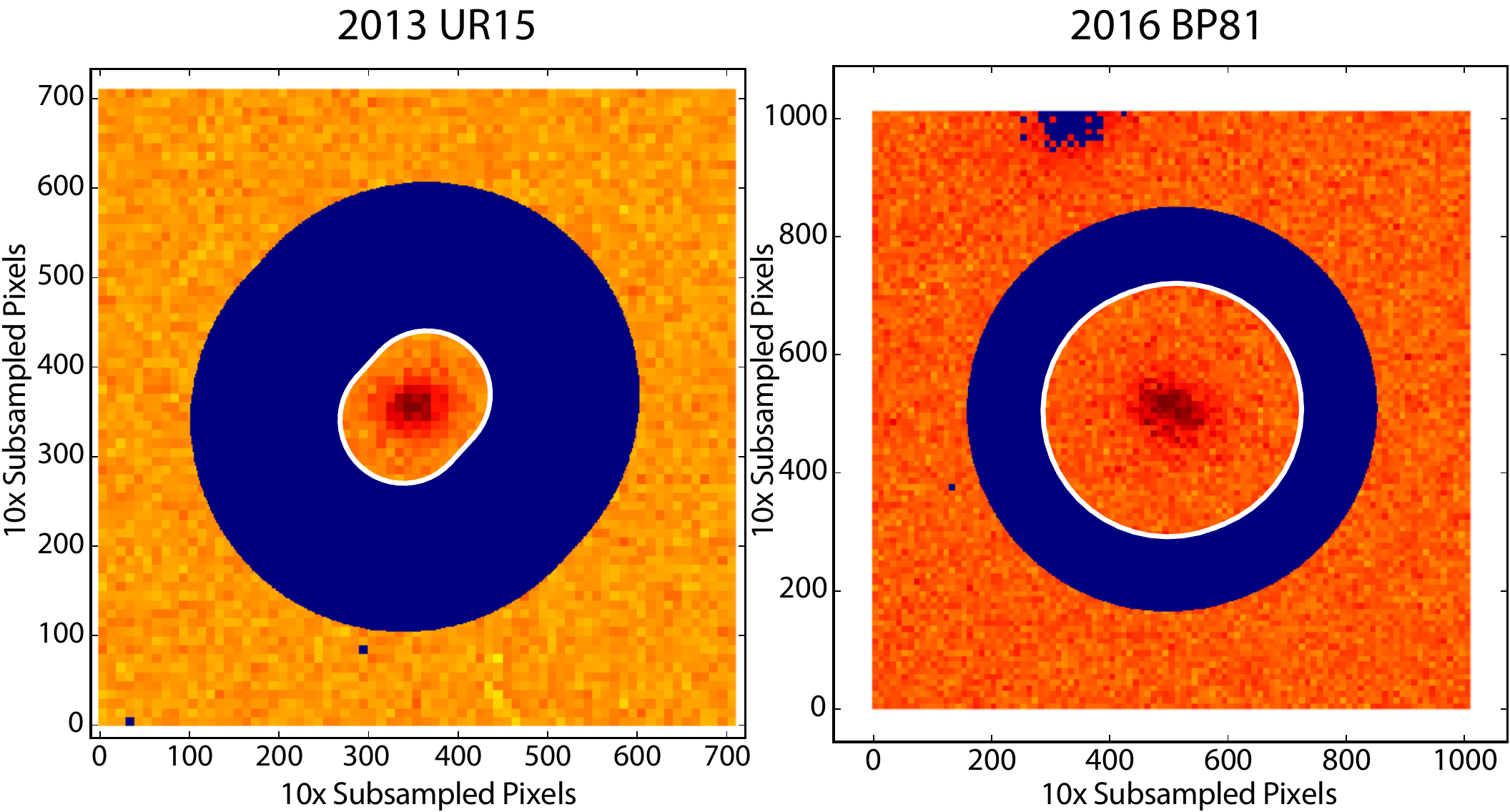}
\caption{\label{fig:optical_data}  Representative GMOS $r$ TRIPPy photometric apertures and sky boxes. The pill apertures are traced in white. Backgrounds were measured outside pill apertures of radii 4 FWHM and inside a square box of width 12 FWHM.  Blue colored pixels are specifically ignored from the analysis. 2013 UR15 is consistent with a trailed point source.  TNO 2016 BP81 is an extended source; it is one of the resolved binaries detected in the Col-OSSOS sample.}
\end{center}
\end{figure*}

Col-OSSOS targets move up to several pixels across the detector during each GMOS integration. Thus, to preserve photometric SNR and avoid the use of unnecessarily large circular apertures, our measurements were made with TRIPPy (Trailed Image Photometry in Python), a dedicated software for photometry of linearly trailed sources \citep{Fraser2016}. Using TRIPPy, we model the image PSF with a 10$\times$ undersampled look up table and a best-fit Moffat profile. PSFs were generated individually for each GMOS frame. Only well isolated stars with photometric $\mbox{SNR}>200$ were used in the PSF generation. In rare frames where the stellar background was particularly sparse, the SNR requirement was reduced (never lower than SNR=80) until at least 3 stars were available. Each star was manually inspected and only those without faint contamination within $\sim$6 Full-Widths at Half Maxima (FWHM) were used. Photometric aperture corrections were measured from the generated PSF. The TRIPPy PSFs were then convolved with a line, with its trail length and angle equal to that of the trailed TNO, to create trailed PSFs (TSFs). These TSFs were then used to compute an optimized pill-shape aperture for our target. Associated pill-shaped  apertures with radii 1.2 FWHM and length equal to the trail length within the image were used to measure source fluxes. 
All pill fluxes were then corrected to 4 FWHM apertures using curves of growth estimated from the TSFs. Backgrounds were measured outside pill apertures of radii 4 FWHM and inside a square box of width 12 FWHM. 
Figure \ref{fig:optical_data} shows an example TRIPPy pill aperture and sky box for representative r-band observations of two Col-OSSOS targets. 
For our reported uncertainties, we adopt the quadrature sum of the photometric shot-noise, the uncertainty in color term, on-image calibration uncertainty, and the uncertainty in curve of growth estimate, which we take as 0.01~mag. 
The photometric uncertainty is dominated by the photometric shot-noise, though we adopt a cautious 0.01 magnitude uncertainty on the aperture corrections. 

An alternate choice for pill photometry would be to use PSF photometry directly. While valid if the inferred PSF and TSFs are accurate reflection of the true image shapes, this is not always the case, even with TRIPPy TSFs. As demonstrated in \citet{Fraser2016}, in the TSF cores, the peak pixels in the TSF can deviate by as much as $\sim8\%$ compared to the true source. While a significantly smaller deviation compared to other photometry packages, this could still  drive the inferred fluxes to be incorrect by a similar amount when using PSF photometry. Aperture photometry, which depends mainly on knowledge of the more correctly modeled TSF wings, is significantly less sensitive to such effects, and as such, allows more accurate flux measurements.

\subsection{Photometric Calibration to Sloan Digital Sky Survey Photometric System}
\label{sec:colorterm}

The GMOS $g$, $r$, and $z$ bandpasses are similar to the filters used by SDSS \citep{York2000}, thus we choose to report our measured colors in the more widely used SDSS Photometric System \citep{Fukugita1996, Padmanabhan2008}. To convert from Gemini magnitudes to SDSS ($g_{\textrm S}$, $r_{\textrm S}$, $z_{\textrm S}$), linear color conversions between the two systems were evaluated using in-frame background stars cataloged in the SDSS \citep{York2000} data release 13 \citep{SDSS2016}, with $g_{\textrm S}<$21 and $r_{\textrm{S}}<$21, $0.3< (g_{\textrm S}-r_{\textrm S})< 1.5$  to span the full range of TNO colors, while avoiding non-linearities in the color correction which occurs at both redder and bluer values. Our approach to measuring color terms uses a least-squares solution over all GMOS stars matched to good SDSS catalog stars in each acquired GMOS image.  Our technique makes use of all Col-OSSOS observations taken to date with the GMOS-N  e2v CCDs (Col-OSSOS observations spanning 2014B-2016B semesters\footnote{Gemini observing programs:
GN-2014B-LP-1,
GN-2015A-LP-1,
GN-2015B-LP-1,
GN-2016A-LP-1,
GN-2016B-LP-1}). All reported Col-OSSOS optical observations taken with the e2v detectors make use of our derived color terms. 

Here we describe the fitting process for the Gemini $r_G$ filter. We apply the same technique to obtain a transformation for the other filters. Circular instrumental aperture fluxes of all SDSS stars observed in GMOS frames were first acquired by TRIPPy.
On image $j$, star $i$ has a $r$ magnitude in the Gemini filter system ($ r_{i,G}$) as:
 \begin{align}
 r_{i,G}  = Z_{j}(r_G) - 2.5 \log{f_{i,j}(r_G)}
 \label{eq:gemini}
\end{align}
where $f_{i,j}(r_G)$ and $Z_{j}(r_G)$ are the instrumental flux and zero point in the GMOS $r$ filter.
The same star's magnitude can also be described by:
\begin{align}
r_{i,G} = r_{i,S} + C\left(g_{i,S}-r_{i,S}\right)
 \label{eq:sloan}
\end{align}
where $r_{i,S}$ is the $r$  band magnitude for star $i$ in the SDSS filter set and $C$ is a linear color term.
Equations \ref{eq:gemini} and \ref{eq:sloan} can be combined as:
\begin{equation}
r_{i,S}+2.5\log{f_{i,j}(r_G)} =  Z_{j}(r_G)-C\left(g_{i,S}-r_{i,S}\right)
 \label{eq:color}
\end{equation}
For the $n$ stars on image $j$, Equation \ref{eq:color} can be written in matrix form as:
\begin{align}
{\left[ \begin{array}{c}
r_{1,S} + 2.5\log{f_{1,j}(r_G)}  \\
\vdots \\
r_{n,S} + 2.5\log{f_{n,j}(r_G)}  \\
\end{array} \right]
= }
\left[ \begin{array}{cc}
1 & -\left(g_{1,S}-r_{1,S}\right) \\
\vdots & \vdots\\
1 & -\left(g_{n,S}-r_{n,S}\right) \\
\end{array} \right]
\left[ \begin{array}{c}
 Z_{j}(r_G) \\
C \\
\end{array} \right]
\end{align}
and by extension, for all $N$ images
\begin{eqnarray}
{\left[ \begin{array}{c}
r_{1,S} + 2.5\log{f_{1,1}(r_G)} \\
r_{2,S} + 2.5\log{f_{2,j}(r_G)} \\
\vdots\\
r_{n,S} + 2.5\log{f_{n,N}(r_G)} \\
\end{array} \right]
= }  \nonumber \\
\left[ \begin{array}{ccccc}
1 & 0 & \dots & 0 & -\left(g_{1,S}-r_{1,S}\right) \\
0 & 1 & \dots & 0 & -\left(g_{2,S}-r_{2,S}\right) \\
\vdots & \vdots & \ddots & \vdots & \vdots \\
0 & 0 & \dots & 1 & -\left(g_{n,S}-r_{n,S}\right) \\
\end{array} \right]
\left[ \begin{array}{c}
Z_1(r_G) \\
Z_2(r_G) \\
Z_j (r_G)\\
\vdots\\
Z_N(r_G) \\
C
\end{array} \right]
\label{sloaneq}
\end{eqnarray}
or $\vec{y}=A\vec{x}$.
Solving Equation \ref{sloaneq},  $\vec{x}=\left(A^{T} \cdot A\right)^{-1} \cdot A^{T} \cdot \vec{y}$ provides the least squares solution for all zero points $Z_j(r_G)$ and color term $C$. 

Uncertainty on $C$ was found by a Monte Carlo process. The magnitude of each star $i$  was scattered by a Gaussian variate distribution with width equal to its photometric uncertainty in the GMOS frames, and a new value $C'$ was found. This process was repeated 200 times, with the uncertainty on $C$ taken as the standard deviation of the set of $C'$.  In total, 754 stars over 102 frames were used to determine the transformation from Gemini to the SDSS system for $r$. For $g$, 753 stars in 188 exposures were used in the least squares fitting. $z$ had the least number of frames and sources used in the fit;  250 SDSS sources over 39 frames were employed in our $z$ color term analysis. The minimum/maximum number of stars per GMOS image used in this analysis was 3/15 for $g$, 2/12 for $r$, and 3/11 for $z$.


From the above analysis, the resultant color terms between the SDSS filter system and the GMOS-e2v filter system are:

\begin{align}
	g_\textrm{G} = g_\textrm{S}  -0.139(\pm0.002)\cdot (g-r)_\textrm{S}\\
	r_\textrm{G} = r_\textrm{S}  -0.060(\pm0.003)\cdot (g-r)_\textrm{S}\\
	z_\textrm{G} = z_\textrm{S}  -0.026(\pm0.017)\cdot (g-r)_\textrm{S}
	\label{eq:colourTerms}
\end{align}

\begin{figure}
\begin{center}
\includegraphics[width=0.7\columnwidth]{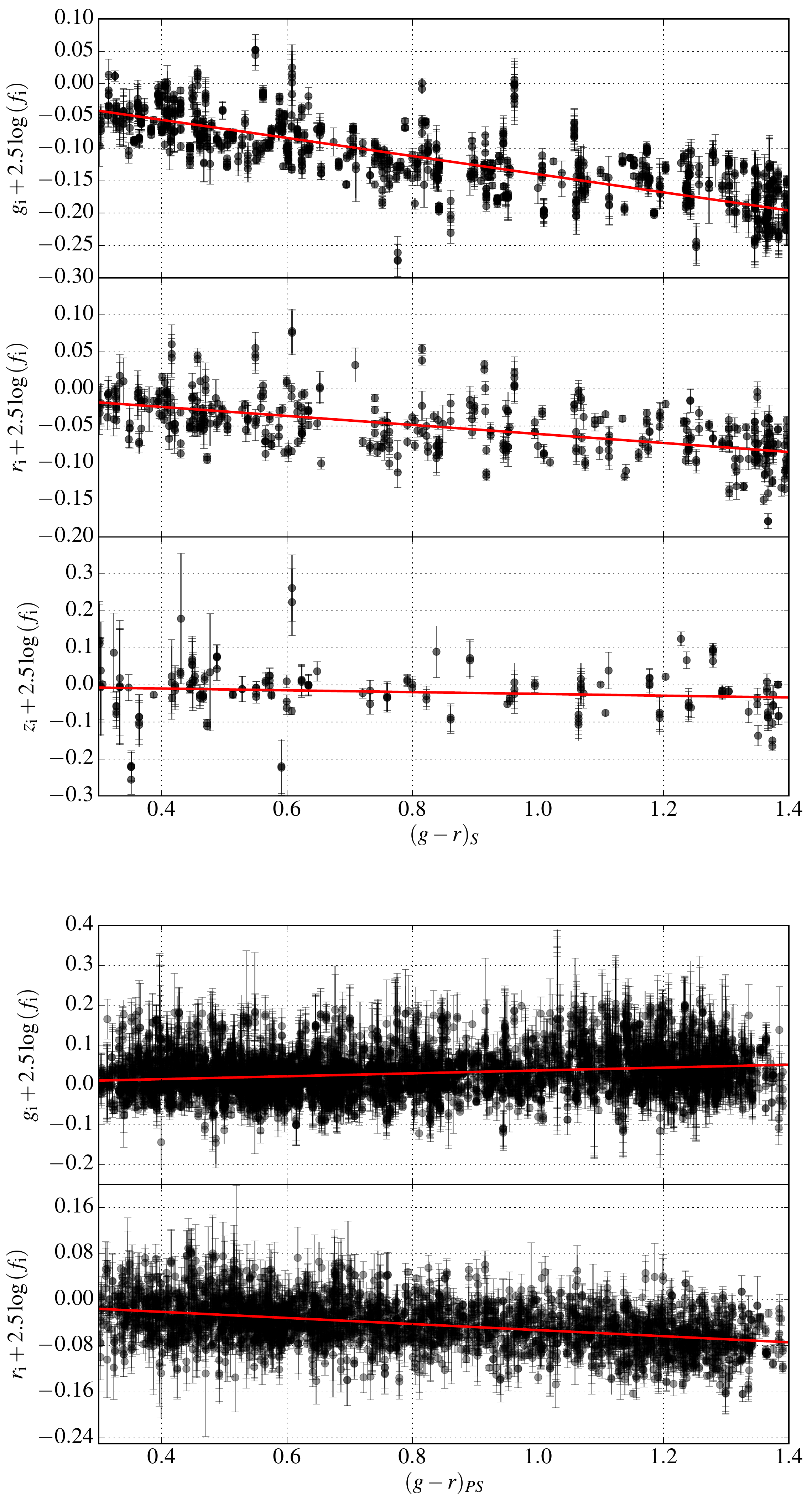}
\caption{\label{fig:sdsscolor} Difference between instrumental Gemini magnitudes and the catalog magnitudes of selected stars, versus catalog color, for the two catalogues used for calibration in Col-OSSOS: SDSS (upper) and Pan-STARRS1 (lower). Upper, SDSS: top: $g$-band, middle: $r$-band, bottom: $z$-band. Lower, Pan-STARRS1: top: $g$-band, bottom: $r$-band. In each case, the derived best-fit linear color term is shown as a red line.}
\end{center}
\end{figure}

\noindent
The color terms to convert to the SDSS photometric system and the best least squares fits are presented in Figure  \ref{fig:sdsscolor}. We note that solving the inverse problem, e.g.  $r_{i,S} = Z_{j}(r_G) - 2.5 \log{f_{i,j}(g_G)} +D \left(g_{\emph{G}}-r_{\emph{G}}\right)$,  results in exactly the same color terms $D$ as those derived from inverting Equation \ref{eq:color}, to within the uncertainties of the color terms $C$. We also note that for consistency between observations, and to ensure we could convert observed $r-z$ colors into the standard SDSS system, we made use of $g-r$ in calculating the $z$ color term. 

For cases where the science frames did not overlap the SDSS fields, we made use of the Pan-STARRS Data Release 1 catalog \citep{2016arXiv161205560C,2016arXiv161205243F, 2016arXiv161205242M}. The linear color conversions between the Pan-STARRS and filter systems was evaluated as described above, but using only stellar-like calibration sources, chosen as those sources with Kron magnitudes and circular aperture magnitudes differing by less than 0.3 mags. The color terms as evaluated against the Pan-STARRS system are given by:

\begin{align}
	g_\textrm{G} = g_\textrm{PS}  +0.037(\pm0.002)\cdot (g-r)_\textrm{PS}\\
	r_\textrm{G} = r_\textrm{PS}  -0.052(\pm0.002)\cdot (g-r)_\textrm{PS}
	\label{eq:colourTermsPS}
\end{align}

\noindent
The color terms to convert to the Pan-STARRS1 system and the best least squares fits are presented in Figure  \ref{fig:sdsscolor}. As a check of the precision of our color terms, we made use of our evaluated color terms to solve for a conversion of $g-r$ between the SDSS and Pan-STARRS system directly, and compared that to the accurate conversion between the two systems reported by \citet{Tonry2012}. Through the valid range in color, $0.3\leq (g-r)_{\textrm{S}} \leq 1.5$, our conversion and that reported by \citet{Tonry2012} deviated by no more than 0.01 magnitudes, demonstrating the veracity of the color terms we report.

The zero points ($Z_j(r_G)$) (which were initially calibrated off the SDSS or Pan-STARRS magnitudes, assuming C=0) in the Gemini system are improved, by converting all cataloged magnitudes of the stars observed in a frame to the GMOS filter system using the derived color terms. The final zero point for image $j$,$Z_{j}(r_G)$, is calculated as the 3-$\sigma$ clipped  weighted mean of all $Z_{j,i}(r_G)$, weighted by the inverse of the quadrature sum of SDSS/Pan-STARRS catalog uncertainty and GMOS photometric uncertainty for each star. The minimum/maximum number of stars per GMOS image used to calculate the final zero points was 11/30 for $g$, 12/29 for $r$, and 10/26 for $z$.

\section{Near-Infrared Data Reduction, Photometry, and Calibration}
\label{sec:NIR}

In this Section, we detail the processing and analysis of the NIRI TNO and standard star observations, and describe our photometric calibration of the NIRI data. After using the Gemini IRAF task nprepare \citep{2016ascl.soft08006G} to prepare the raw image data and add keyword information to the FITS (Flexible Image Transport System) headers, we employed custom built python scripts to reduce the images and measure fluxes.

\subsection{NIRI Data Reduction}
\label{sec:NIR_reduction}

All, NIRI science and calibrator images were passed  through cosmic ray and bad pixel rejection routines. Cosmic ray rejection was performed on each NIRI frame using the python implementation\footnote{\url{https://obswww.unige.ch/\~tewes/cosmics\_dot\_py}} of the L.A.Cosmic (Laplacian Cosmic Ray Identification) algorithm \citep{vanDokkum2001}. Bad pixel maps were created by using both the individual dark exposures and co-added nightly dark exposures to identify outlier pixels, those with unusual sensitivity behavior.  Bright/hot pixels are identified from the short exposure dark as those pixels that are 2.5 times standard deviation above the median value of the dark. Low-sensitivity pixels are those that are 4.5 sigma below the median of the dark and as pixels with negative values. We then combined our bad pixel map with the one provided by the Gemini IRAF package. These flagged pixels were not used when combining the individual NIRI frames to produce the stacks. Additionally, science frames that showed a mean source flux variation of more than 50\% compared to the mean source fluxes of the entire sequence were flagged as likely suffering from extinction due to cloud or moisture and not included in the creation of the sky frames and the final image stacks.

For a given Col-OSSOS target sequence, a master sky flat was produced from the median unshifted science frames with sources masked. In order to account for temporal variations in the sky, a custom sky frame was generated uniquely for each NIRI TNO image taken. The sky frame was produced from a rolling average of the 15 temporally closest usable images with sources masked, and avoiding common dither patters, with the window temporally centered around the frame, when ever possible. We settled on 15 images for the sky frame after examining the frame-to-frame variation produced by differing numbers of images, as fewer frames produced a background measurement of insufficient quality from the noisy individual images. This custom sky frame was scaled to match the background level of the science frame in question, and then subtracted. In our first release sample (described in Section \ref{sec:sample}), the minimum number of NIRI frames in a TNO sequence is 7 (2010 RE188) and the maximum was 79 (2013 UN15). 

The individual sky-subtracted TNO images were then mean combined in two ways:  (1) a sidereal stack and (2) a non-sidereal stack. The sidereal stack  is made by co-adding the individual frames using multiple star centroids in order to create a deep image of the star field and obtain point-like stars that can be used to compute the PSF.  The non-sidereal stack is created by co-adding the frames shifting at the TNO's predicted on-sky velocity, as determined from propagating the OSSOS best-fit orbit to the time of observation, using the OSSOS python wrapper\footnote{Available at \url{https://github.com/OSSOS/liborbfit}, and from the Python Package Index via \texttt{pip install mp\_ephem}} of the \citet{Bernstein2000} orbit fitter. For each TNO target, the full NIRI sequence was divided at the sequence mid point into two half sequences, from which two sidereal and two non-sidereal stacks (called split stacks) were produced. This afforded some sensitivity to light curve variations, which we discuss in Section \ref{sec:colours}. When the number of frames to be stacked was less than 9, a median stack was produced so as to afford some resistance to errant pixels when a small number of frames was available. Otherwise a mean stack was produced.  The minimum number of NIRI frames used in a split stack was 3 and the maximum was 40. The effective exposure time for each stack and the number of NIRI frames associated with each non-sidereal stack is listed in Table \ref{tab:sample_obs}. 

NIR photometric standard stars bracketed the full GMOS-NIRI-GMOS TNO sequence. For each standard, the frames were aligned and median-combined to produce sidereal calibration stacks from which the standard star's flux could be measured. The standard star observations were divided by a flat field produced from the GCAL (Gemini facility calibration unit) imaging flats obtained as part of the daily calibrations, resulting in images with flat background, exhibiting brightness variations of less than 1\% across the whole frame. 

\subsection{Near-Infrared Photometry and Photometric Calibration}
\label{sec:NIR_calibration}
\begin{figure*}
\begin{center}
\includegraphics[width=0.9\textwidth]{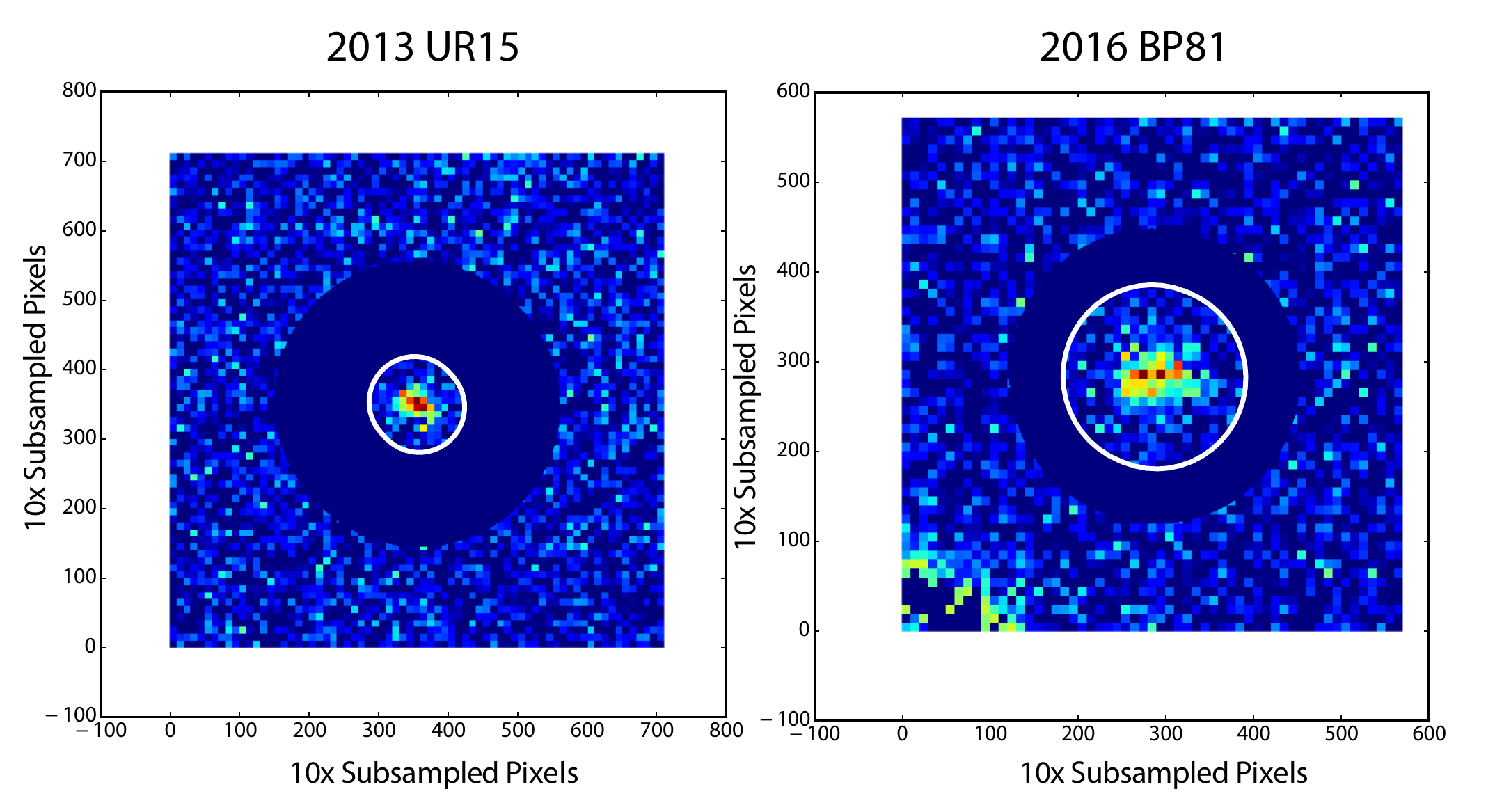}
\caption{\label{fig:NIRI_data}   Two representative NIRI $J$ non-sidereal stacks with the Col-OSSOS target's TRIPPy photometric apertures and sky boxs plotted. The blue colored annulii are specifically ignored from the analysis. 2013 UR15 is consistent with a trailed point source. TNO 2016 BP81 is an extended source; it is one of the resolved binaries found in the Col-OSSOS sample.}
\end{center}
\end{figure*}

For the reasons discussed in Section \ref{sec:optical_phot}, we adopt pill apertures to measure photometry from our NIRI observations. The PSF was generated from the sidereal stacks.  The relatively low number of non-saturated stars, due to NIRI's small FOV, and the presence of barely resolved galaxies made it challenging to generate an accurate PSF. Particular care was therefore taken in choosing the stars used to generate the PSF. Where possible, hand-selected point-like stars with $SNR>100$ were used to generate the stellar PSFs, and the TSF of each individual source. In a few cases where at least three satisfactory stars were unavailable, this SNR threshold was lowered as necessary to include at least three stars in the PSF generation, to a lower limit of SNR=50. In one case, only 2 suitable stars were available for PSF generation, regardless of the SNR threshold.
 Appropriate pill apertures tailored to each TNO target with radius $1.2$~FWHM, and curve of growth corrections, measured from the TSF, were then used to measure the flux of the TNOs in the non-sidereal stacks. We note that the alignment necessary to produce the sidereal and non-sidereal stacks may have induced small variations in the true stacked TSF that are not properly reflected in the TSF generated by convolution of the PSF, which was generated from the sidereal stack. This would reflect in the aperture correction derived for the non-sidereal stacks. Experiments in removing the TSF profile from the science frames produced residuals of only a few percent. Thus, for our reported photometry we adopt a generous 0.02 magnitude uncertainty (double that of the GMOS value) which reflects this additional level of complication. Sky backgrounds were measured outside pill apertures of radii 4 FWHM and inside a square box of width 12 FWHM.  Figure \ref{fig:NIRI_data} shows the TRIPPy photometric aperture and sky box for the non-sidereal stacks of two representative Col-OSSOS targets.

 We calculate each target's NIR colors based on the mean $J$-band magnitudes derived from the two split stacks. The consistency of our $J$-band photometry is demonstrated by the close match between the mean measurement derived from the two split stacks, and the measurement derived from the stack built from the full sequence of images. Specifically, for all the TNOs in our first-release dataset (Section \ref{sec:sample}), both measurements agree within 0.03 mag, in agreement with our photometric precision. There is only one exception, 2010 RE188, where the two measurements agree within 0.06 mag. 

The photometric uncertainty for each of the non-sidereal stacks was taken as the square root of the quadratic sum of all of the error contributions. Four main sources of uncertainty were identified in our near-infrared photometric measurements: the photometric shot-noise, accuracy of the science stack background level measurement, zero point measurement error, and the uncertainty due to the TSF estimation. For the background estimation error, we adopt a value of 0.02~mag for stacks with fewer than 15 images, and a value of 0.01~mag otherwise. The uncertainty on the aperture correction was calculated as the standard deviation in aperture corrections measured from each of the sources used to generate the TSF and PSF. We adopted a minimum value of  0.01 magnitudes if the PSF was computed from at least 3 stars and 0.02 for 2 stars. Finally, we include the 0.02 magnitude uncertainty as a result of the uncertain aperture correction. We find that the resultant SNR performance is in agreement with expectations from the NIRI performance (reported by the Gemini Integration Calculator\footnote{\url{http://www.gemini.edu/sciops/instruments/integration-time-calculators/niri-itc}}) for the range of target brightnesses considered here.

Photometric calibration was preformed using the standard stars observation that bracket the full NIRI-GMOS sequence. Using TRIPPy, a large 4~FWHM radius circular aperture was used to measure the flux of each calibrator, and thus infer a zero point for the calibration frame. The zero points measured in each calibrator stack were corrected to reflect the mean airmass and precipitable water vapor reported by the Gemini weather monitors for each TNO science stack we generated. The adopted zero point for each TNO stack was taken as the mean of the two calibration measurements acquired. 

\section{Appendix: Full Photometry Sequences and Photometric Variability of the OSSOS 13BL and 14BH TNO Sample}
\label{sec:aphotplots}

\begin{figure*}
\begin{center}
\includegraphics[width=0.90\textwidth]{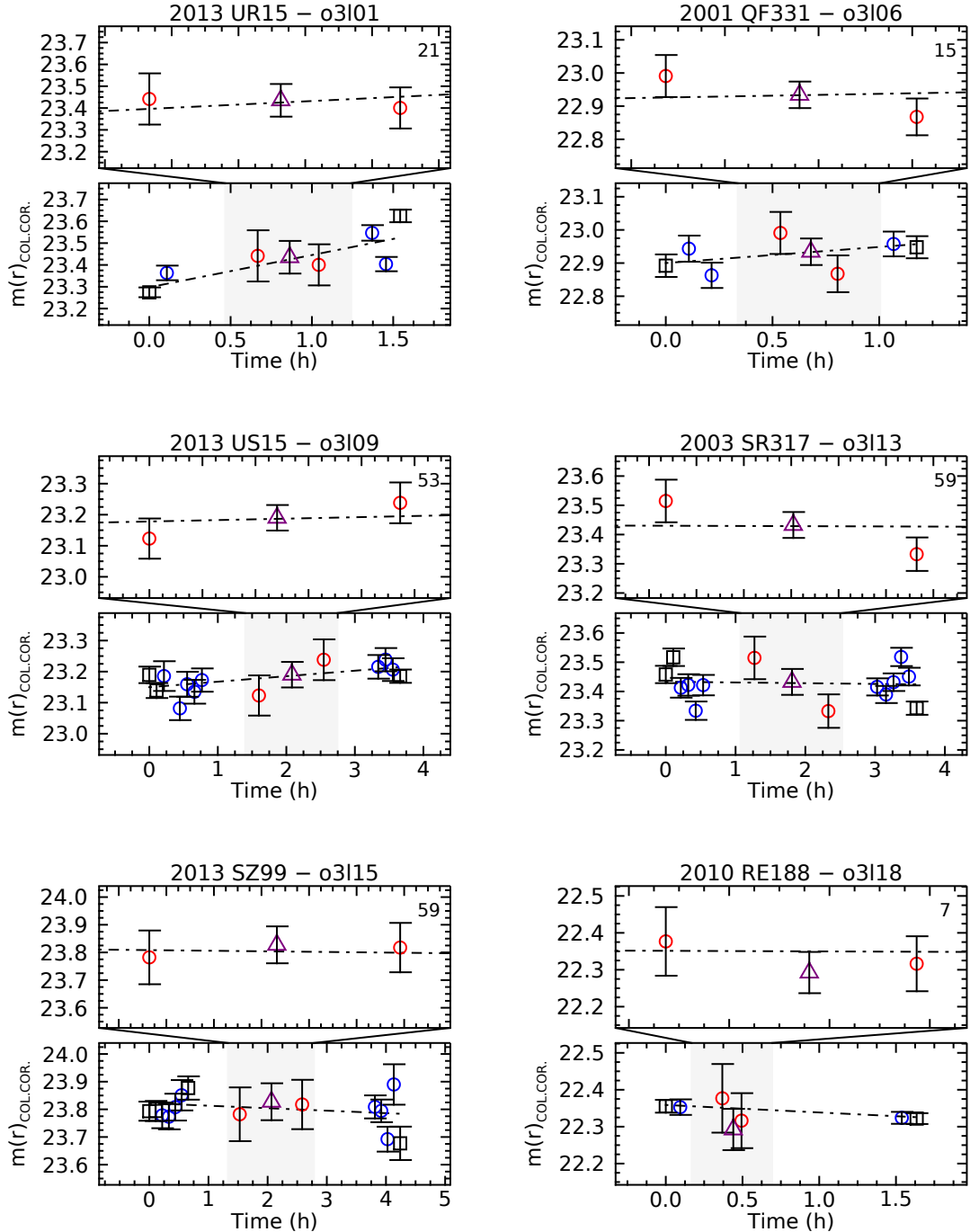}
\caption{\label{fig:photometry_plotted_1} Col-OSSOS optical and near-infrared observed photometry and variability of the OSSOS 13BL and 14BH TNO sample. Each panel is divided into two sub-panels showing the full GMOS/NIRI $(z)rgJgr(z)$ photometric sequence (bottom) and a zoom on the $J$-band measurements alone (top). Each measurement was color-corrected to a $r$-band magnitude using the color terms derived for the corresponding object. All optical magnitudes are in the SDSS photometric system and $J$-band magnitudes are in the MKO system. {\it Black squares:} $r$-band magnitudes, {\it blue circles}: color-corrected $g$-band magnitudes, {\it orange circles}: color-corrected $z$-band magnitudes, {\it red circles}: color-corrected $J$-band split-stack magnitudes, {\it purple triangles}: color-corrected $J$-band full-stack magnitudes. The number of near-infrared frames acquired for each object is indicated in the upper right corner of each panel. The best-fit light curve for each target is shown by the dashed line. Plots for all TNOs presented in this paper are available in the online supplemental data.}
\end{center}
\end{figure*}

We present in the online supplemental figures  the measured photometry in each Col-OSSOS GMOS/NIRI optical-NIR-optical sequence plotted for each target TNO in the 13BL and 14BH OSSOS blocks. For each sequence, the fitted linear light curve used to derive the target's photometric colors, as described in Section \ref{sec:colours}, is also plotted. Figure \ref{fig:photometry_plotted_1} is presented as a representative example. 

\facilities{Gemini:Gillett (GMOSN, NIRI), CFHT:MegaCam}
\software{astropy \citep{Astropy2013, Astropy2018}, 
Gemini IRAF package \citep{2016ascl.soft08006G}, 
IRAF \citep{Tody1986},
L.A.Cosmic \citep{vanDokkum2001} in python\footnote{Malte Tewes, 2010 \url{http://obswww.unige.ch/~tewes/cosmics_dot_py/cosmics.py_0.4/doc/index.html}}, 
matplotlib \citep{Hunter2007}, 
mp\_ephem\footnote{\url{https://github.com/OSSOS/liborbfit} }
MegaPipe \citep{2008PASP..120..212G},
NumPy \citep{Oliphant2006}, 
PyRAF, 
pysynphot software package \citep{Lim+2015}, 
SExtractor \citep{Bertin1996},  
SciPy \citep{Jones2001},  
synphot,
TRIPPy \citep{Fraser2016}
}

\bibliographystyle{apj}

\end{CJK*}
\end{document}